\documentclass[10.5]{article}

\usepackage{mypackage} 
\usepackage{mycommand} 

\algnewcommand{\algorithmicforeach}{\textbf{for each}}
\algdef{SE}[FOR]{ForEach}{EndForEach}[1]
  {\algorithmicforeach\ #1\ \algorithmicdo}
  {\algorithmicend\ \algorithmicforeach}

\usepackage[pdfencoding=auto,hypertexnames=false]{hyperref}
\hypersetup{colorlinks=false}

\numberwithin{equation}{section} 
\usepackage{mytheoremeng} 

\title{Optimal ancilla-free Clifford+$T$ synthesis \\ for general single-qubit unitaries}

\author[1]{Hayata Morisaki \thanks{u748119d@ecs.osaka-u.ac.jp}}
\author[2]{Kaoru Sano}
\author[2,3]{Seiseki Akibue \thanks{seiseki.akibue@ntt.com}}

\affil[1]{Graduate School of Engineering Science, The University of Osaka \\
 1-3 Machikaneyama, Toyonaka, Osaka 560-8531, Japan.}
\affil[2]{NTT Institute for Fundamental Mathematics, NTT Communication Science Labs., NTT, Inc.\\
3--1, Morinosato-Wakamiya, Atsugi, Kanagawa 243-0198, Japan}
\affil[3]{NTT Research Center for Theoretical Quantum Information, NTT, Inc.\\
3--1, Morinosato-Wakamiya, Atsugi, Kanagawa 243-0198, Japan.}

\date{}

\begin{document}
\maketitle

\vspace{-5mm}
\begin{abstract}
    We propose two Clifford+$T$ synthesis algorithms that are optimal with respect to $T$-count. 
    The first algorithm, called \emph{deterministic synthesis}, approximates any single-qubit unitary by a single-qubit Clifford+$T$ circuit with the minimum $T$-count. 
    The second algorithm, called \emph{probabilistic synthesis}, approximates any single-qubit unitary by a probabilistic mixture of single-qubit Clifford+$T$ circuits with the minimum $T$-count. 
    For most of single-qubit unitaries, the runtimes of deterministic synthesis and probabilistic synthesis are $\varepsilon^{-1/2 - o(1)}$ and $\varepsilon^{-1/4 - o(1)}$, respectively, for an approximation error $\varepsilon$. 
    Although this complexity is exponential in the input size, we demonstrate that our algorithms run in practical time at $\varepsilon \approx 10^{-15}$ and $\varepsilon \approx 10^{-22}$, respectively. 
    Furthermore, we show that, for most single-qubit unitaries, the deterministic synthesis algorithm requires at most $3\log_2(1/\varepsilon) + o(\log_2(1/\varepsilon))$ $T$-gates, and the probabilistic synthesis algorithm requires at most $1.5\log_2(1/\varepsilon) + o(\log_2(1/\varepsilon))$ $T$-gates. 
    Remarkably, complexity analyses in this work do not rely on any numerical or number-theoretic conjectures.
\end{abstract}

\section{Introduction}

The set of implementable quantum gates is limited in fault-tolerant quantum computation (FTQC). 
Therefore, approximating a given target unitary, within a desired approximation error $\varepsilon$ by using only such a gate set is an important problem, which is referred to as the \emph{synthesis problem}. 
In particular, the Clifford+$T$ gate set is currently implementable in many FTQC schemes\cite{knill1998resilient,knill1998resilient2,litinski2018game,thxx-njr6,hirano2024leveraging,gidney2024magic}, and its synthesis problem has been extensively studied. 
In such FTQC schemes, the implementation cost of the $T$ gate is often significantly higher than that of Clifford gates, and the $T$-count is commonly used as a measure of circuit complexity. 

A wide variety of algorithms have been proposed for solving the synthesis problem.
The Solovay-Kitaev algorithm~\cite{dawson2005solovay} approximates single-qubit unitaries by using $\mathcal{O}\bigl(\log^{3.97} (1/\varepsilon)\bigr)$ gates from a fixed universal gate set and runs in $\text{polylog}(1/\varepsilon)$ time. 
Because it exploits the group-theoretic structure of the unitary group, it can be applied to any universal gate set.
However, the number of gates it produces is far larger than the optimum.
On the other hand, recent work has shifted toward number-theoretic synthesis algorithms that exploit the algebraic properties of the specific gate set under consideration. 
By leveraging these properties, the gate-count minimization problem can be recast as an algebraic optimization task, yielding circuits with dramatically fewer gates.
Kliuchnikov \textit{et al}.~\cite{kliuchnikov2013asymptotically} proposed a polynomial time Clifford+$T$ synthesis algorithm for single-qubit unitaries that uses only $\mathcal{O}(\log (1/\varepsilon))$ $T$ gates and a constant number of ancilla qubits. 
In 2016, Ross and Selinger~\cite{ross2016optimal} introduced a polynomial time Clifford+$T$ synthesis algorithm for $z$-rotations that uses only $\mathcal{O}(\log (1/\varepsilon))$ $T$ gates and requires no ancilla qubits. 
Using this synthesis algorithm, Kliuchnikov \textit{et al}.~\cite{kliuchnikov2023shorter} proposed an ancilla-free algorithm that approximates single-qubit unitaries with a near-optimal $T$-count (heuristically about $7\log_2(1/\varepsilon)$) and runs in polynomial time. 
However, no practical and optimal synthesis algorithm with respect to $T$-count is known to date. 

Also, to go beyond the limitations of the deterministic setting, it is necessary to consider \emph{probabilistic synthesis}—a recent technique that approximates a unitary channel by a mixed unitary channel. 
For example, \cite{PhysRevA.95.042306} proposes a probabilistic synthesis algorithm for $z$-rotations, achieving about half the $T$-count of the $T$-optimal deterministic Clifford+$T$ synthesis for $z$-rotations~\cite{ross2016optimal}. 
However, because probabilistic synthesis requires additional computational costs to determine the probability distribution and the list of unitary channels, it remains nontrivial whether one can design an optimal probabilistic synthesis algorithm with a practical runtime.

The runtime and $T$-count of existing synthesis algorithms are typically estimated based on numerical or number-theoretic assumptions. Providing precise guarantees for the runtime and $T$-count enables a more rigorous estimation of the resources required for FTQC algorithms. 
Moreover, it offers a new direction for the theoretical analysis of future work on synthesis algorithms.

\subsection{Our results}

In this work, we consider two synthesis problems defined below.
\begin{prob}
\label{prob: synthesis problem}
    Given a single-qubit unitary $V$ and an approximation error $\varepsilon > 0$, we define the two ancilla-free Clifford+$T$ synthesis problems for single-qubit unitaries as follows:
    \begin{itemize}
        \item \textbf{$T$-optimal Deterministic Synthesis} \\
        Find a single-qubit Clifford+$T$ circuit $U$ with minimum $T$-count such that 
        $
            \frac12\| \mathcal{U} - \mathcal{V} \|_\diamond < \varepsilon.
        $ 

        \item \textbf{$T$-optimal Probabilistic Synthesis} \\
        Find a finite set of single-qubit Clifford+$T$ circuits $\{U_x\}_{x \in X}$ and a probability distribution $p$ on $X$ with minimum $\max_{x \in X} \mathcal{T}(U_x)$ such that
        $
            \frac12\left\| \sum_{x \in X} p(x)\mathcal{U}_x - \mathcal{V} \right\|_\diamond < \varepsilon.
        $
    \end{itemize}
    Here, $\mathcal{U}$ and $\mathcal{V}$ denote the quantum channels corresponding to the unitary operators $U$ and $V$, and $\mathcal{T}(U)$ denotes the $T$-count of a Clifford+$T$ circuit $U$. 
\end{prob}

We propose algorithms to solve these two problems and provide a theoretical analysis of their performance based on measure (probability) theory, without any conjectures. 
The following two theorems summarize the discussions presented in Sections~\ref{sec: Deterministic Synthesis} and \ref{sec: Probabilistic Synthesis}, respectively.
\begin{thm}[Informal Summary of Section~\ref{sec: Deterministic Synthesis}]
    There exists an algorithm that solves $T$-optimal Deterministic Synthesis.
    For most of single-qubit unitaries, the algorithm runs in time $\varepsilon^{-1/2 - o(1)}$ and produces circuits whose $T$-count is $3\log_2(1/\varepsilon) + o(\log_2(1/\varepsilon))$. 
\end{thm}

\begin{thm}[Informal Summary of Section~\ref{sec: Probabilistic Synthesis}]
    There exists an algorithm that solves $T$-optimal Probabilistic Synthesis.
    For most of single-qubit unitaries, the algorithm runs in time $\varepsilon^{-1/4 - o(1)}$ and produces circuits whose $T$-count is $1.5\log_2(1/\varepsilon) + o(\log_2(1/\varepsilon))$. 
\end{thm}

In Section~\ref{sec: Numerical Experiment}, we present a numerical performance evaluation of our proposed synthesis algorithms. 
These results show that the deterministic and probabilistic synthesis algorithms are feasible up to $\varepsilon \approx 10^{-15}$ and $\varepsilon \approx 10^{-22}$, respectively. 
Moreover, within these approximation errors, we verify that all 100 Haar-random target unitaries used in our experiments can be approximated with about $3\log_2(1/\varepsilon)$ $T$-gates in the deterministic setting and about $1.5\log_2(1/\varepsilon)$ $T$-gates in the probabilistic setting.

\vspace{-1.5mm}
\begin{table}[h]
    \renewcommand{\arraystretch}{2}
    \setlength{\tabcolsep}{3pt}
    \centering
    \begin{tabular}{|c|c|c||c|c|}
        \hline
        & \multicolumn{2}{c||}{\textbf{Deterministic Synthesis}} 
        & \multicolumn{2}{c|}{\textbf{Probabilistic Synthesis}} \\
        \hline
        \textbf{Algorithms} 
        & \makecell[c]{\textbf{$T$-count} \\ \textbf{estimate}} & \makecell[c]{\textbf{Runtime} \\ \textbf{estimate}} 
        & \makecell[c]{\textbf{$T$-count} \\ \textbf{estimate}} & \makecell[c]{\textbf{Runtime} \\ \textbf{estimate}}  \\
        \hline\hline
        \makecell[c]{ RS method base~\cite{ross2016optimal, PhysRevA.95.042306} \\ (Near $T$-optimal) } 
        & $9 \log_{2}(1/\varepsilon)$ \textsuperscript{*}
        & polylog$(1/\varepsilon)$  \textsuperscript{*}
        & $4.5 \log_{2}(1/\varepsilon)$ \textsuperscript{*}
        & polylog$(1/\varepsilon)$  \textsuperscript{*} \\
        \hline
        
        \makecell[c]{ State-of-the-art~\cite{kliuchnikov2023shorter} \\ (Near $T$-optimal) } 
        & $7 \log_{2}(1/\varepsilon)$ \textsuperscript{*}
        & polylog$(1/\varepsilon)$  \textsuperscript{*}
        & $3.5 \log_{2}(1/\varepsilon)$ \textsuperscript{*}
        & polylog$(1/\varepsilon)$  \textsuperscript{*} \\
        \hline
        
        \makecell[c]{Brute-force~\cite{fowler2011constructing} \\ ($T$-optimal)} 
        & $3 \log_{2}(1/\varepsilon) $ \textsuperscript{*}
        & \makecell[c]{$\widetilde{O}(\varepsilon^{-3})$ \textsuperscript{*}  \\ (feasible up to $\varepsilon \approx 10^{-4}$) }  
        & \multicolumn{2}{c|}{} \\
        \hline
        
        \makecell[c]{\textbf{This work}~\\ ($T$-optimal)} 
        & \makecell[c]{$3 \log_2(1/\varepsilon)$ \\[1mm] for most of $\U(2)$} 
        &  \makecell[c]{$\varepsilon^{-1/2-o(1)}$ \\[1mm] for most of $\U(2)$ \\[1mm]  (feasible up to $\varepsilon \approx 10^{-15}$)  } 
        & \makecell[c]{$1.5 \log_{2}(1/\varepsilon)$ \\[1mm] for most of $\U(2)$} 
        & \makecell[c]{$\varepsilon^{-1/4-o(1)}$ \\[1mm] for most of $\U(2)$  \\[1mm] (feasible up to $\varepsilon \approx 10^{-22}$) } \\
        \hline
    \end{tabular}

    \vspace{-1mm}
    \caption{Comparison of ancilla-free Clifford+$T$ synthesis algorithm for single-qubit unitaries. Those marked with * denote heuristic complexity.}
    \label{tab:comparison}
\end{table}

\section{Preliminary \label{sec: preliminary}}

\subsection{Basic notations}
In this chapter, we introduce the basic notation and mathematical properties that will be used throughout this paper. 

First, we define the linear-algebraic notation used in quantum computation. 
This work focuses exclusively on finite-dimensional complex Euclidean spaces $\C^n$; other cases are not considered. 
We define the sets of complex matrices as follows:
\begin{itemize}
    \item $\Mat(n) \coloneqq \C^{n \times n}$, $n\times n$ complex matrices,
    \item $\Herm(n) \coloneqq \{ A \in \Mat(n) \ | \ A = A^\dag \}$, $n$-dimensional hermitian matrices,
    \item $\Pos(n) \coloneqq \{ A \in \Herm(n) \ | \ \forall x \in \C^n, \ x^\dag A x \geq 0 \}$, $n$-dimensional positive semidefinite matrices,
    \item $\U(n) \coloneqq \{ A \in \Mat(n) \ | \ A^\dag A = I_n \}$, $n$-dimensional unitary group,
    \item $\SU(n) \coloneqq \{ U \in \U(n) \ | \ \det(U) = 1 \}$, $n$-dimensional special unitary group,
    \item $\PU(n) \coloneqq \U(n) / \U(1) \cong \SU(n) / (\Z/n)$ , $n$-dimensional projective unitary group,
    \item $\StateSpace(n) \coloneqq \{ \rho \in \Pos(n) \ | \ \tr[\rho] = 1 \}$, $n$-dimensional quantum states,
    \item $\PureStateSpace(n) \coloneqq \{ \rho \in \StateSpace(n) \ | \ \tr[\rho^2] = 1 \}$, $n$-dimensional quantum pure states,
\end{itemize}
where $I_n$ is the $n$-dimensional identity matrix. 
For $A, B \in \Herm(n)$, we write $A \leq B$ to mean that $B - A \in \Pos(n)$. 
Next, we define the sets of all quantum channels from $\StateSpace(n)$ to $\StateSpace(m)$ by $\CPTP(n,m)$, and the set of all $n$-dimensional unitary channels by 
\begin{equation}
    \UC(n) \coloneqq \{ \mathrm{Ad}_{U} \ | \ U \in \U(n) \} \quad \text{where $\mathrm{Ad}_U(\rho) \coloneqq U \rho U^\dag$ $\bigl(\forall \rho \in \Mat(n) \bigr)$}.
\end{equation}
In particular, we refer to $\UC(2)$ as the \emph{single-qubit unitary group} and call each of its elements a \emph{single-qubit unitary}. 
Let $\Phi : \PU(n) \rightarrow \UC(n)$ be the mapping well-defined by
\begin{equation}
    \Phi([U]) \coloneqq \text{Ad}_U.
\end{equation}
It is well known that $\Phi$ is a topological group isomorphism---that is, $\UC(n) \cong \PU(n) = \U(n) / \U(1)$. 
Accordingly, we adopt the following convention throughout this paper.
\begin{conv*}
    We denote unitary matrices by capital letters (e.g., $U, V, W$) and the corresponding unitary channels by calligraphic letters (e.g., $\mathcal{U}, \mathcal{V}, \mathcal{W}$), where $\mathcal{U}(\rho) = U \rho U^\dag$. 
    Also, for convenience of exposition, we often identify a unitary channel $\mathcal{U}$ with its implementing unitary matrix $U$. 
    Since $\text{Ad}_U = \text{Ad}_{e^{i\theta}U}$ for any $\theta \in \R$, the choice of global phase remains ambiguous. 
    However, with our discussion, this identification is independent of the choice of global phase. 
\end{conv*}

Next, we introduce the diamond distance, the metric used throughout this paper to quantify the closeness of two quantum channels. 
Let $\Phi : \Mat(n) \rightarrow \Mat(m)$ be a linear mapping. 
The diamond norm of $\Phi$ is defined by
\begin{equation}
    \| \Phi\|_\diamond \coloneqq \max_{\rho \in \StateSpace(n^2)} \| (\Phi \otimes \mathcal{I}_n)(\rho)  \|_{\tr},
\end{equation}
where $\mathcal{I}_n$ is the identity operator on $\Mat(n)$ and $\| M \|_{\tr} = \tr[\sqrt{M^\dag M}]$. 
Then, the diamond distance is the distance induced by the norm; for $\mathcal{A}, \mathcal{B} \in \CPTP(n,m)$, we define the diamond distance of $\mathcal{A}$ and $\mathcal{B}$ by
\begin{equation}
\label{eq: dfn of diamond distance}
    d_\diamond(\mathcal{A}, \mathcal{B}) \coloneqq \frac12 \| \mathcal{A} - \mathcal{B} \|_\diamond. 
\end{equation}
The diamond distance possesses several properties that are useful for evaluating the closeness between two quantum algorithms (i.e., quantum channels) and is widely used as a metric \cite{Watrous_2018}. 
Among these properties, we will frequently use the bi-invariance of the diamond distance for unitary channels---for all $\mathcal{U}, \mathcal{V}, \mathcal{W} \in \UC(n)$, it holds that
\begin{equation}
    d_\diamond(\mathcal{U}, \mathcal{V}) = d_\diamond(\mathcal{U}\mathcal{W}, \mathcal{V}\mathcal{W}) = d_\diamond(\mathcal{W}\mathcal{U}, \mathcal{W}\mathcal{V}). 
\end{equation}
As shown in \cite{Watrous_2018}, the value of Eq.~\eqref{eq: dfn of diamond distance} equals the optimal value of the following semidefinite programming (SDP): 
\begin{equation}
\label{eq: SDP compute diamond distance}
    \begin{tabular}{rlcrl}
    \multicolumn{2}{c}{\underline{{\rm Primal problem}}} &\ \ \ \ \ \ \ \ \ \ \ \ \ \  
     &\multicolumn{2}{c}{\underline{{\rm Dual problem}}}\\
    {\rm maximize:}&$\tr[J(\mathcal{A}-\mathcal{B})T]$&&{\rm minimize:}&$r\in\mathbb{R}$ \\
    {\rm subject to:}& $0\leq T\leq\rho\otimes I_m$,&&
    {\rm subject to:}& $S\geq0\wedge S\geq J(\mathcal{A}-\mathcal{B})$,\\
    &$\rho \in \mathbf{S}(n).$&&&$rI_n\geq\tr_{\C^m}[S]$.\\
    \end{tabular} 
\end{equation}
where $J(\Phi) = \sum_{1 \leq i,j \leq n} \ket{i}\bra{j} \otimes \Phi(\ket{i}\bra{j})$ is the Choi-Jamio\l{}kowski representation of $\Phi : \Herm(n) \rightarrow \Herm(m)$. 
In general, computing the diamond distance is difficult to express in a closed analytical form. 
However, this is not the case for single-qubit unitaries. 
\begin{prop}[Diamond distance between single-qubit unitaries]
\label{prop: diamond distance between single-qubit unitaries}
    Let $U$ and $V$ be single-qubit unitaries. 
    It holds that
    \begin{equation}
    \label{eq: diamond distance between signle-qubit unitaries}
        \diamonddist{\mathcal{U}}{\mathcal{V}} = \sqrt{1 - \tfrac14\bigl|\tr[UV^\dag]\bigr|^2}. 
    \end{equation}
\end{prop}
\begin{proof}
    Let $W = UV\dag$, and let $e_1, e_2 \in \C$ be the eigenvalues of $W$. 
    Form Theorem~B.1 of \cite{kliuchnikov2023shorter}, we have
    \begin{equation}
        \diamonddist{\mathcal{U}}{\mathcal{V}} = \frac12|e_1 - e_2|.
    \end{equation}
    Since $W \in \U(2)$, there exist $w_1, w_2 \in \C$ and $\theta \in \R$ such that
    \begin{equation}
        W = e^{i\theta}\left(\begin{array}{rr}
            w_1 & -w_2^\dag \\
            w_2 &  w_1^\dag
        \end{array}\right). 
    \end{equation}
    Then, the eigenvalues $e_1, e_2$ satisfy the following equation:
    \begin{equation}
        \lambda^2 - 2e^{i\theta} \Re(w_1)\lambda + e^{2i\theta} = 0.
    \end{equation}
    Therefore, we obtain
    \begin{equation}
        |e_1 - e_2| = \left| \sqrt{4\Re(w_1)^2 - 4} \right| = 2\sqrt{1 - \tfrac14\left|\tr[W]\right|^2}. 
    \end{equation}
\end{proof}

Next, we describe our notations for measure theory and probability theory. 
Let $(\Omega, \mathcal{F}, \mu)$ be a probability space and $(E, \mathcal{E})$ be a measurable space. 
For a random variable (measurable function) $X : \Omega \rightarrow E$ and a proposition $P$ with respect to $X(\omega)$, if there exists $A \in \mathcal{E}$ such that $P \Longleftrightarrow X(\omega) \in A$ for all $\omega \in \Omega$, we define the probability of $P$ as
\begin{equation}
    \Pr_{\omega \sim \mu} \left[ P \right] \coloneqq \mu\bigl( X^{-1}(A) \bigr). 
\end{equation}
Also, if $E = \R$ and $X$ is an integrable function, we denote the expectation value of $X$ by
\begin{equation}
    \Expect_{\mu} \left[ X \right] \coloneqq \int_{\Omega} X(\omega) d\mu(\omega). 
\end{equation}
For a locally compact topological group $G$, we denote the probability (normalized) Haar measure on $G$ by $\mu_G$. 
Throughout this paper, we consider probability spaces only with respect to $\mu_{\UC(2)}$ or the Haar measure on a locally compact topological group isomorphic to $\UC(2)$. 
Therefore, to simplify notation, we write $\Pr$ instead of $\Pr\limits_{V \sim \mu_\UC(2)}$ and $\mathbb{E}$ instead of $\mathbb{E}_{\mu_\UC(2)}$. 
Next, we define the set of probability distributions over a finite set $X$ as
\begin{equation}
    \ProbDist(X) \coloneqq \left\{ p : X \rightarrow [0,1] \ \middle| \ \sum_{x \in X} p(x) = 1 \right\}.
\end{equation}

We define the open ball of center $U \in \UC(2)$ and radius $r > 0$ in the metric space $(\UC(2), \ d_\diamond)$ by
\begin{equation}
    \Ball{r}{U} \coloneqq \left\{ V \in \UC(2) \ | \ \diamonddist{\mathcal{U}}{\mathcal{V}} < r \right\}. 
\end{equation}
It is known that, for all $U \in \UC(2)$, $\Pr[V \in \Ball{\varepsilon}{U}] = \mu_{\UC(2)}\bigl(\Ball{\varepsilon}{U}\bigr) = \Theta(\varepsilon^3)$ as $1/\varepsilon \rightarrow \infty$. 
Also, for a finite set of $\{ U_x \in \UC(2) \}_{x \in X}$, a subset $S \subseteq \UC(2)$, and a positive real $\delta > 0$, we define
\begin{equation}
    \text{$\{U_x\}_{x\in X}$ is a $\delta$-covering of $S$} 
    \overset{\text{def}}{\iff} 
     S \subseteq \bigcup_{x \in X} \Ball{\delta}{U_x}
\end{equation}

\subsection{Asymptotic notations}
In this chapter, we introduce the notion of \textit{with high probability}. 
Let $(\Omega, \mu, \mathcal{F})$ be a probability space. 
For an event $E_n$ depending on such a parameter $n$, we say that
\begin{center}
    $E_n$ holds with high probability (w.h.p.) as $n \rightarrow \infty$
\end{center}
if it holds that
\begin{equation}
\label{eq: dfn of with high probability}
    \Pr_{\omega \sim \mu}[E_n] = 1 - O\left( n^{-c} \right) \quad \text{as $n \rightarrow \infty$}
\end{equation}
for some $c > 0$ independent of $n$. 
Next, we apply this notation to the big-$O$ and small-$o$ notations. 
Let $X_n, Y_n : \Omega \rightarrow \R$ be random variables depending on such a parameter $n$, and let $f$ be a real function. 
We define the following asymptotic notations:
\begin{align}
    \label{eq: dfn X_n = Owhp(f(n))}
    &X_n = \Owhp(f(n)) \ \text{as $n \rightarrow \infty$} 
    \ \overset{\mathrm{def}}{\Longleftrightarrow} \ 
    \exists C > 0,\; \text{$|X_n(\omega)| \leq C|f(n)|$ holds w.h.p. as $n \rightarrow \infty$},\ and \\
    \label{eq: dfn X_n = owhp(f(n))}
    &X_n = \owhp(f(n)) \ \text{as $n \rightarrow \infty$} 
    \ \overset{\mathrm{def}}{\Longleftrightarrow} \ 
    \forall \varepsilon > 0,\; \text{$|X_n(\omega)| \leq \varepsilon|f(n)|$ holds w.h.p. as $n \rightarrow \infty$}.
\end{align}
Next, we define the asymptotic notation between random variables:
\begin{align}
    \label{eq: dfn X_n = Owhp(Y_n)}
    &X_n = \Owhp\bigl(Y_n\bigr) \ \text{as $n \rightarrow \infty$} \ \overset{\mathrm{def}}{\Longleftrightarrow} \
    \exists C > 0 ,\; \text{$|X_n(\omega)| \leq C|Y_n(\omega)|$ holds w.h.p. as $n \rightarrow \infty$}, \ and \\
    \label{eq: dfn X_n = owhp(Y_n)}
    &X_n = \owhp\bigl(Y_n\bigr) \ \text{as $n \rightarrow \infty$} \ \overset{\mathrm{def}}{\Longleftrightarrow} \
    \forall \varepsilon > 0 ,\; \text{$|X_n(\omega)| \leq \varepsilon|Y_n(\omega)|$ holds w.h.p. as $n \rightarrow \infty$}. 
\end{align}
In particular, if $X_n(\omega)$ is asymptotically bounded by $Y_n(\omega)$ for all $\omega \in \Omega$, we write as follows:
\begin{align}
    \label{eq: dfn X_n = O(Y_n)}
    &X_n = O\bigl(Y_n\bigr) \ \text{as $n \rightarrow \infty$} \ \overset{\mathrm{def}}{\Longleftrightarrow} \ \exists C > 0,\; \exists n_C > 0,\; |X_n(\omega)| \leq C|Y_n(\omega)| \ \text{for all $n \geq n_C$ and $\omega \in \Omega$},  \\
    \label{eq: dfn X_n = o(Y_n)}
    &X_n = o\bigl(Y_n\bigr) \ \text{as $n \rightarrow \infty$} \ \overset{\mathrm{def}}{\Longleftrightarrow} \ \forall \varepsilon > 0,\; \exists n_\varepsilon > 0,\; |X_n(\omega)| \leq \varepsilon|Y_n(\omega)| \ \text{for all $n \geq n_\varepsilon$ and $\omega \in \Omega$}. 
\end{align}
Next, we introduce some properties of the asymptotic notation $\Owhp$ used in this work. 
Note that by taking $Y_n(\omega) = f(n)$, which is a constant function with respect to $\omega$, the definition in Eq~\eqref{eq: dfn X_n = Owhp(f(n))} is included in the definition in Eq~\eqref{eq: dfn X_n = Owhp(Y_n)}. 
\begin{prop}[Properties of $\Owhp$]
\label{prop: properties of Owhp}
    All asymptotic analyses in this lemma are taken as $n \rightarrow \infty$; therefore, this condition is not stated explicitly. 
    The following equations hold: 
    \begin{enumerate}[label=\textbf{\arabic*.}]
        \item $X_n = O(Y_n)$ $\Longrightarrow$ $X_n = \Owhp(Y_n)$

        \item $X_n = \Owhp(Y_n)$ and $Y_n = \Owhp(Z_n)$ $\Longrightarrow$ $X_n = \Owhp(Z_n)$ 

        \item $X_n^{(1)} = \Owhp(Y_n^{(1)})$ and $X_n^{(2)} = \Owhp(Y_n^{(2)})$ $\Longrightarrow$ $X_n^{(1)} + X_n^{(2)} = \Owhp\left(\bigl|Y_n^{(1)}\bigr| + \bigl|Y_n^{(2)}\bigr| \right)$ 

        \item $X_n^{(1)} = \Owhp(Y_n^{(1)})$ and $X_n^{(2)} = \Owhp(Y_n^{(2)})$ $\Longrightarrow$ $X_n^{(1)} X_n^{(2)} = \Owhp(Y_n^{(1)} Y_n^{(2)})$

        \item $X_n = \Owhp(Y_n)$ $\Longrightarrow$ $\forall c > 0,\; |X_n|^c = \Owhp(|Y_n|^c)$

    \end{enumerate}
\end{prop}

\begin{proof}
    We show each proof as follows.
    \begin{enumerate}[label=\textbf{\arabic*.}]    
        \item It is trivial by definition. 

        \item There exist $C_1, C_2 > 0$ such that $\Pr_{\omega \sim \mu}[|X_n(\omega)| > C_1 |Y_n(\omega)|] = O(\poly(1/n))$ and $\Pr_{\omega \sim \mu}[|Y_n(\omega)| > C_2|Z_n(\omega)|] = O(\poly(1/n))$. It holds that 
        \begin{align}
            \Pr_{\omega \sim \mu}[|X_n(\omega)| > C_1 C_2 |Z_n(\omega)|] &\leq \Pr_{\omega \sim \mu}[|X_n(\omega)| > C_1|Y_n(\omega)| \ \vee \ |Y_n(\omega)| > C_2|Z_n(\omega)|] \\
            &\leq \Pr_{\omega \sim \mu}\left[|X_n(\omega)| > C_1|Y_n(\omega)| \right] + \Pr_{\omega \sim \mu}\left[|Y_n(\omega)| > C_2|Z_n(\omega)| \right] \\
            &= O(\poly(1/n)).
        \end{align}

        \item There exist $C_1, C_2 > 0$ such that $\Pr_{\omega \sim \mu}[|X_n^{(1)}(\omega)| > C_1 |Y_n^{(1)}(\omega)|] = O(\poly(1/n))$ and $\Pr_{\omega \sim \mu}[|X_n^{(2)}(\omega)| > C_2|Y_n^{(2)}(\omega)|] = O(\poly(1/n))$. It holds that
        \begin{align}
            &\Pr_{\omega \sim \mu}\left[ \bigl|X_n^{(1)}(\omega) + X_n^{(2)}(\omega)\bigr| > \max(C_1, C_2)\left( \bigl|Y_n^{(1)}(\omega)\bigr| + \bigl|Y_n^{(2)}(\omega)\bigr| \right)  \right] \\ 
            \leq &\Pr_{\omega \sim \mu}\left[ \bigl|X_n^{(1)}(\omega) \bigr| > \max(C_1, C_2)\bigl| Y_n^{(1)}(\omega) \bigr| \vee \bigl|X_n^{(2)}(\omega) \bigr| > \max(C_1, C_2)\bigl| Y_n^{(2)}(\omega) \bigr| \right] \\
            \leq &\Pr_{\omega \sim \mu}\left[ \bigl| X_n^{(1)}(\omega) \bigr| > \max(C_1, C_2)\bigl| Y_n^{(1)}(\omega) \bigr| \right] + \Pr_{\omega \sim \mu}\left[ \bigl| X_n^{(2)}(\omega) \bigr| > \max(C_1, C_2)\bigl| Y_n^{(2)}(\omega) \bigr| \right] \\
            = &O(\poly(1/n)).
        \end{align}

        \item There exist $C_1, C_2 > 0$ such that $\Pr_{\omega \sim \mu}[|X_n^{(1)}(\omega)| > C_1 |Y_n^{(1)}(\omega)|] = O(\poly(1/n))$ and $\Pr_{\omega \sim \mu}[|X_n^{(2)}(\omega)| > C_2|Y_n^{(2)}(\omega)|] = O(\poly(1/n))$. It holds that
        \begin{align}
            &\Pr_{\omega \sim \mu}\left[ \left|X_n^{(1)}(\omega) X_n^{(2)}(\omega) \right| > C_1 C_2\left| Y_n^{(1)}(\omega) Y_n^{(2)}(\omega) \right| \right] \\
            \leq &\Pr_{\omega \sim \mu}\left[ \bigl|X_n^{(1)}(\omega) \bigr| > C_1\bigl| Y_n^{(1)}(\omega) \bigr| \vee \bigl|X_n^{(2)}(\omega) \bigr| > C_2\bigl| Y_n^{(2)}(\omega) \bigr| \right] \\
            \leq &\Pr_{\omega \sim \mu}\left[ \bigl| X_n^{(1)}(\omega) \bigr| > C_1\bigl| Y_n^{(1)}(\omega) \bigr| \right] + \Pr_{\omega \sim \mu}\left[ \bigl| X_n^{(2)}(\omega) \bigr| > C_2\bigl| Y_n^{(2)}(\omega) \bigr| \right] \\
            = &O(\poly(1/n)). 
        \end{align}

        \item There exists $C > 0$ such that $\Pr_{\omega \sim \mu}\left[ |X_n(\omega)| > C|Y_n(\omega)| \right] = O(\poly(1/n))$. It holds that
        \begin{align}
            \Pr_{\omega \sim \mu}\left[ |X_n(\omega)|^c > C^c|Y_n(\omega)|^c \right] &= \Pr_{\omega \sim \mu}\left[ |X_n(\omega)| > C|Y_n(\omega)| \right] \\
            &= O(\poly(1/n)).
        \end{align}
    \end{enumerate}
\end{proof}

We consider several limit variables. 
For a real function $g$, we introduce the following notation:
\begin{align}
    \text{$E_n$ is holds w.h.p. as $g(n) \rightarrow \infty$} 
    \ \overset{\text{def}}{\Longleftrightarrow} \ 
    \exists c > 0, \; \Pr_{\omega \sim \mu}\left[ E_n \right] = 1 - O\left( g(n)^{-c} \right) \ \text{as $g(n) \rightarrow \infty$}. 
\end{align}
Also, this asymptotic notation is naturally extended to the notations defined in Eqs.~\eqref{eq: dfn X_n = Owhp(f(n))}-\eqref{eq: dfn X_n = o(Y_n)}. 
Proposition~\ref{prop: properties of Owhp} holds for these asymptotic notations.

\subsection{Number-theoretical properties of Clifford+$T$ gate set \label{subsec: number-theoretical properties}}
First, we introduce the basic notation used throughout this paper for the Clifford+$T$ gate set. 
The single-qubit Clifford group $\mathcal{C}_1$ is defined as the group generated by the following two single-qubit unitaries
\begin{equation}
    H = \frac{1}{\sqrt{2}} \begin{pmatrix}
        1 & 1 \\
        1 &-1 
    \end{pmatrix} \quad \text{and} \quad
    S = \begin{pmatrix}
        1 & 0 \\
        0 & i 
    \end{pmatrix}.
\end{equation}
The single-qubit Clifford+$T$ gates, denoted $\CliffordTOne$, is the group generated by $\mathcal{C}_1$ and
\begin{equation}
    T = \begin{pmatrix}
        1 & 0 \\
        0 & e^{i \pi/4}
    \end{pmatrix}.
\end{equation} 
For $U \in \CliffordTOne$, the $T$-count of $U$, denoted $\Tcount(U)$, is defined by 
\begin{equation}
    \Tcount(U) \coloneqq   \min \Bigl\{
   n\in\mathbb{N}\ \Bigm|\ 
   \exists\,C_0,\dots,C_n\in\mathcal{C}_1, \ 
   C_0\,T\,C_1\,T\,\dots T\,C_n \;=\; U
 \Bigr\}.
\end{equation}
Also, for $t \in \N$, we define two subsets of $\CliffordTOne$ used often in this paper:
\begin{align}
    \CliffordTOne_{\Tcount = t} &\coloneqq \{ U \in \CliffordTOne \ | \ \Tcount(U) = t \}, \ \text{the single-qubit Clifford+$T$ gates with $t$ $T$-gates}, \\
    \CliffordTOne_{\Tcount \leq t} &\coloneqq \{ U \in \CliffordTOne \ | \ \Tcount(U) \leq t \}, \ \text{the single-qubit Clifford+$T$ gates with at most $t$ $T$-gates}.
\end{align}

Recently, several studies \cite{kliuchnikov2013asymptotically, ross2016optimal, kliuchnikov2012fast, selinger2012efficient} have used number theory to analyze the properties of quantum circuits generated by certain fixed finite gate sets. 
To pave the way for the number-theoretic Clifford+$T$ synthesis algorithm presented in Section~\ref{sec: Deterministic Synthesis}, we introduce the number-theoretic notations and facts relevant to this work. 
\begin{dfn}[Extensions of $\Z$]
    Recall that $\zeta = e^{i\pi / 4}$.
    Throughout this paper, we use the following rings:
    \begin{itemize}
        \item $\Zsqrttwo = \{ a + b\sqrt{2} \ | \ a,b \in \Z \}$
        \item $\Zzeta = \left\{ a + b\zeta + c\zeta^2 + d\zeta^3 \ | \ a,b,c,d \in \Z \right\} = \Bigl( \Zsqrttwo + i\Zsqrttwo \Bigr) \cup \Bigl( \Zsqrttwo + i\Zsqrttwo + \zeta \Bigr)$
        \item $\D = \left\{ \frac{a}{2^k} \ | \ a \in \Z, k \in \N \right\}$
        \item $\Dzeta = \left\{ a + b\zeta + c\zeta^2 + d\zeta^3 \ | \ a,b,c,d \in \D \right\}$ 
    \end{itemize}
\end{dfn}

\begin{dfn}[Denominator exponent]
\label{dfn: denominator exponent}
    Let $u \in \Dzeta$ and $k \in \N$. If $\sqrt{2}^k u \in \Zzeta$, then we say that $k$ is a \emph{denominator exponent} of $u$. 
    The smallest such $k$ is the \emph{least denominator exponent} of $u$, denoted by $\lde(u)$. 
    The notion extends to analogously vectors and matrices whose entries are in $\Dzeta$. 
\end{dfn}

Kliuchnikov \textit{et al}.~\cite{kliuchnikov2012fast} showed that the representation matrix of single-qubit Clifford+$T$ operators can be characterized over $\Dzeta$. Moreover, Corollary 7.11 of \cite{giles2013remarks} establishes an explicit relationship between the least denominator exponent and the $T$-count. 
The following theorem unifies these two results.
\begin{thm}[Number theoretic property of $\CliffordTOne$]
\label{thm: exact Clifford+T synthesis}
    Let $U$ be a single-qubit unitary. 
    $U$ belongs to $\CliffordTOne$ if and only if there exist $u_1, u_2 \in \Dzeta$ and $l \in \{0,1\}$ such that
    \begin{equation}
    \label{eq: representation matrix of Clifford+T}
        U = \left( \begin{array}{rr}
            u_1 & -u_2^\dag \zeta^l \\
            u_2 &  u_1^\dag \zeta^l
        \end{array} \right).
    \end{equation}
    Moreover, the $T$-count of $U$ satisfies
    \begin{equation}
    \label{eq: relation t-count and lde}
        \Tcount(U) \in \begin{cases}
            \{ 2k - 2, \ 2k \} & \text{if $l = 0$} \\[4pt]
            \{2k - 3, \ 2k - 1, \ 2k + 1 \} & \text{if $l = 1$} \\
        \end{cases},
    \end{equation}
    where $k = \lde(U)$.
\end{thm}

Moreover, Kliuchnikov \textit{et al}.~\cite{kliuchnikov2012fast} proposed an exact synthesis algorithm that, given a single-qubit unitary in the form of Eq.~\ref{eq: representation matrix of Clifford+T}, finds a single-qubit Clifford+$T$ sequence with minimum $T$-count in $O\bigl( \Tcount(U) \bigr)$ arithmetic operations. 
Using Theorem~\ref{thm: exact Clifford+T synthesis} and this algorithm, we can reduce the Clifford+$T$ synthesis problem to a search problem over $\Dzeta$.

\section{Deterministic Synthesis \label{sec: Deterministic Synthesis}}
In this section, we construct an algorithm to solve the following problem. 
\begin{prob}[$T$-optimal Deterministic Synthesis]
\label{prob: dete synth}
    Given a single-qubit unitary $V \in \UC(2)$ and an approximation error $\varepsilon > 0$, find $U \in \CliffordTOne$ with minimum $T$-count such that \begin{equation}
        \diamonddist{\mathcal{U}}{\mathcal{V}} < \varepsilon .
    \end{equation} 
\end{prob}

Our strategy of $T$-count minimization is straightforward: we iteratively solve the following subproblem for each natural number $t$ in ascending order. 
As shown in \cite{harrow2002efficient}, any single-qubit unitary can be approximated to precision $\varepsilon$ by a single-qubit Clifford+$T$ gate with $T$-count $O(\log(1/\varepsilon))$. 
Consequently, this subproblem needs to be solved for at most $O(\log (1/\varepsilon) )$ distinct values of $t$, so the brute-force minimization of the $T$-count is reasonable. 
\begin{prob}[Fixed $T$-count Deterministic Synthesis]
\label{prob: sub dete synth}
    Given a single-qubit unitary  $V \in \UC(2)$, an approximation error $\varepsilon > 0$ and $t \in \N$, enumerate all $U \in \CliffordTOne_{\Tcount = t}$ such that $\diamonddist{\mathcal{U}}{\mathcal{V}} < \varepsilon$, i.e., the set $\CliffordTOne_{\Tcount = t} \cap \mathbf{B}_\varepsilon(V)$. 
\end{prob}

\subsection{Reduction to integer-point enumeration \label{subsec: reduction to integer point enum}}
We show that Problem~\ref{prob: sub dete synth} can be reduced to an integer-point enumeration problem and, based on this reduction, construct an algorithm to solve it.
From Theorem~\ref{thm: exact Clifford+T synthesis}, we can see that if there exists a solution in Problem~\ref{prob: sub dete synth}, then
\begin{itemize}
    \item if $t$ is even, there exist $u_1, u_2 \in \Dzeta$ such that $|u_1|^2 + |u_2|^2 = 1$, $\lde(U) \leq \frac{t+2}{2}$ and $\diamonddist{\mathcal{U}}{\mathcal{V}} < \varepsilon$, where
    \begin{equation}
    \label{eq: form of U even-t}
        U = \left( \begin{array}{rr}
            u_1 & -u_2^\dag \\
            u_2 &  u_1^\dag
        \end{array} \right),
    \end{equation}
    
    \item if $t$ is odd, there exist $u_1, u_2 \in \Dzeta$ such that $|u_1|^2 + |u_2|^2 = 1$, $\lde(U) \leq \frac{t+3}{2}$ and $\diamonddist{\mathcal{U}}{\mathcal{V}} < \varepsilon$, where
    \begin{equation}
        U = \left(\begin{array}{rr}
            u_1 & -u_2^\dag \zeta \\
            u_2 &  u_1^\dag \zeta
        \end{array}\right).
    \end{equation}
\end{itemize}
We consider solving Problem~\ref{prob: sub dete synth} by finding $u_1, u_2 \in \Dzeta$ that satisfy the above conditions. 
Also, since the bi-invariance of the diamond distance and the fact that $\lde(u) = \lde(\zeta u)$ for all $u \in \Dzeta$, the conditions on $u_1, u_2 \in \Dzeta$ when $t$ is odd can be reformulated as follows:
\begin{itemize}
    \item if $t$ is odd, there exist $u_1, u_2 \in \Dzeta$ such that $|u_1|^2 + |u_2|^2 = 1$, $\lde(U) \leq \frac{t+3}{2}$ and 
    $d_\diamond(\mathcal{U}, \ \mathcal{VT}^\dag) < \varepsilon$
    , where 
    \begin{equation}
    \label{eq: error condition t is odd}
        U = \left(\begin{array}{rr}
            u_1 & -u_2^\dag \\
            u_2 &  u_1^\dag
        \end{array}\right).
    \end{equation}
\end{itemize}
Consequently, when $t$ is odd, redefining the target unitary as $V \leftarrow VT^\dag$ reduces the problem to the same conditions as the even-$t$ case. 
Suppose that $U$ can be written in the form of Eq.~\eqref{eq: form of U even-t} and that
$V = 
    \left( \begin{array}{rr}
       v_1 & -v_2^\dag \\
       v_2 &  v_1^\dag
    \end{array}
    \right)
$, we have 
\begin{equation}
\label{eq: parameter representation of diamond distance U and V}
    \diamonddist{\mathcal{U}}{\mathcal{V}} = \sqrt{1 - \Re\bigl( u_1 v_1^\dag + u_2 v_2^\dag \bigr)^2} 
\end{equation}
by Proposition~\ref{prop: diamond distance between single-qubit unitaries}.
If we identify a two-dimensional complex vector $(z_1, z_2)^\text{t}$ with the four-dimensional real vector
\begin{equation}
    \Vec{z} =  \bigl(\Re(z_1), \ \Im(z_1), \ \Re(z_2), \  \Im(z_2)\bigr)^\text{t},
\end{equation}
then the $\Re\bigl( u_1 v_1^\dag + u_2 v_2^\dag \bigr)$ coincides exactly with the standard Euclidean inner product $\Vec{u} \cdot \Vec{v}$. 
Note that this identification is made via the isomorphism $\UC(2) \cong \SU(2) / (\Z/2) \cong S^3 / (\Z/2)) \eqqcolon \widetilde{S}^3$. 
Hence, $\text{Eq.\eqref{eq: parameter representation of diamond distance U and V}} < \varepsilon$ is equivalent to
\begin{equation}
    |\vec{u} \cdot \vec{v}| > \sqrt{1 - \varepsilon^2}.
\end{equation}
From Eq.~\eqref{eq: diamond distance between signle-qubit unitaries}, it follows that the maximum value of the diamond distance between single-qubit unitaries is $1$. Therefore, when considering the synthesis problem for single-qubit unitaries, we may assume without loss of generality that $\varepsilon \leq 1$. 
Summarily, we can reduce Problem~\ref{prob: sub dete synth} to the following problem.
\begin{prob}
\label{prob: enum u1 and u2 for fixed k}
    Given a target vector 
    $\vec{v} \in \widetilde{S}^3$, an approximation error $\varepsilon \in (0,1]$ and $k \in \N$, 
    find $\vec{u} \in \Dzeta^2$ such that
    \begin{enumerate}
        \item\label{item: lde constraint} $\sqrt{2}^k \vec{u} \in \Zzeta^2$,  
        \item\label{item: norm constraint} $\| \Vec{u} \| = 1$,
        \item\label{item: error constraint} $ |\vec{u} \cdot \vec{v}| > \sqrt{1 - \varepsilon^2}$.
    \end{enumerate}
\end{prob}

Next, we reduce Problem~\ref{prob: enum u1 and u2 for fixed k} to enumerating the integer points contained in a convex body $S \subseteq \R^d$, since an efficient algorithm exists for solving this problem \cite{lenstra1983integer}. 
\begin{prop}[Integer Enumeration Algorithm]
\label{prop: integer enum algorithm}
    Fix a dimension $d$. For any convex body $S \subseteq \R^d$, there exists an algorithm that enumerates the integer points in $S$, i.e., the set $S \cap \Z^d$, using $O\bigl( |S \cap \Z^d| + \log \vol(S) \bigr)$ arithmetic operations. 
\end{prop}
\noindent
In preparation for the reduction, we introduce the automorphism $(-)^\bullet $ on $\Dzeta$, defined by
\begin{equation}
    \bigl(a + b\zeta + c\zeta^2 + d\zeta^3 \bigr)^\bullet = a - b\zeta + c\zeta^2 - d\zeta^3,
\end{equation}
and denote by $\vec{u}^\bullet$ the four-dimensional real vector identified with the $\bigl( u_1^\bullet, u_2^\bullet \bigr)^\text{t} \in \Dzeta^2$. 
\begin{lem}
    Let $\vec{u} \in \Dzeta$ be a solution to Problem~\ref{prob: enum u1 and u2 for fixed k}. 
    Then the following conditions hold:
    \begin{align}
        \label{eq: u in A}
        \vec{u} &\in \Bigl\{ \vec{x} \in \R^4 \ \Bigm| \ \|\Vec{x}\| \leq 1 , \ |\Vec{x} \cdot \Vec{v}| > \sqrt{1 - \varepsilon^2}  \Bigr\} \eqqcolon \mathcal{R}_\varepsilon(\vec{v}) \quad and, \\[4pt]
        \label{eq: u^dot in B}
        \vec{u}^\bullet &\in \Bigl\{ \Vec{x} \in \R^4 \ \Bigm| \ \|\vec{x}\| \leq 1 \Bigr\} \eqqcolon \mathcal{D}. 
    \end{align}
\end{lem}

\begin{figure}[h]
    \centering
    \includegraphics[width=1\linewidth]{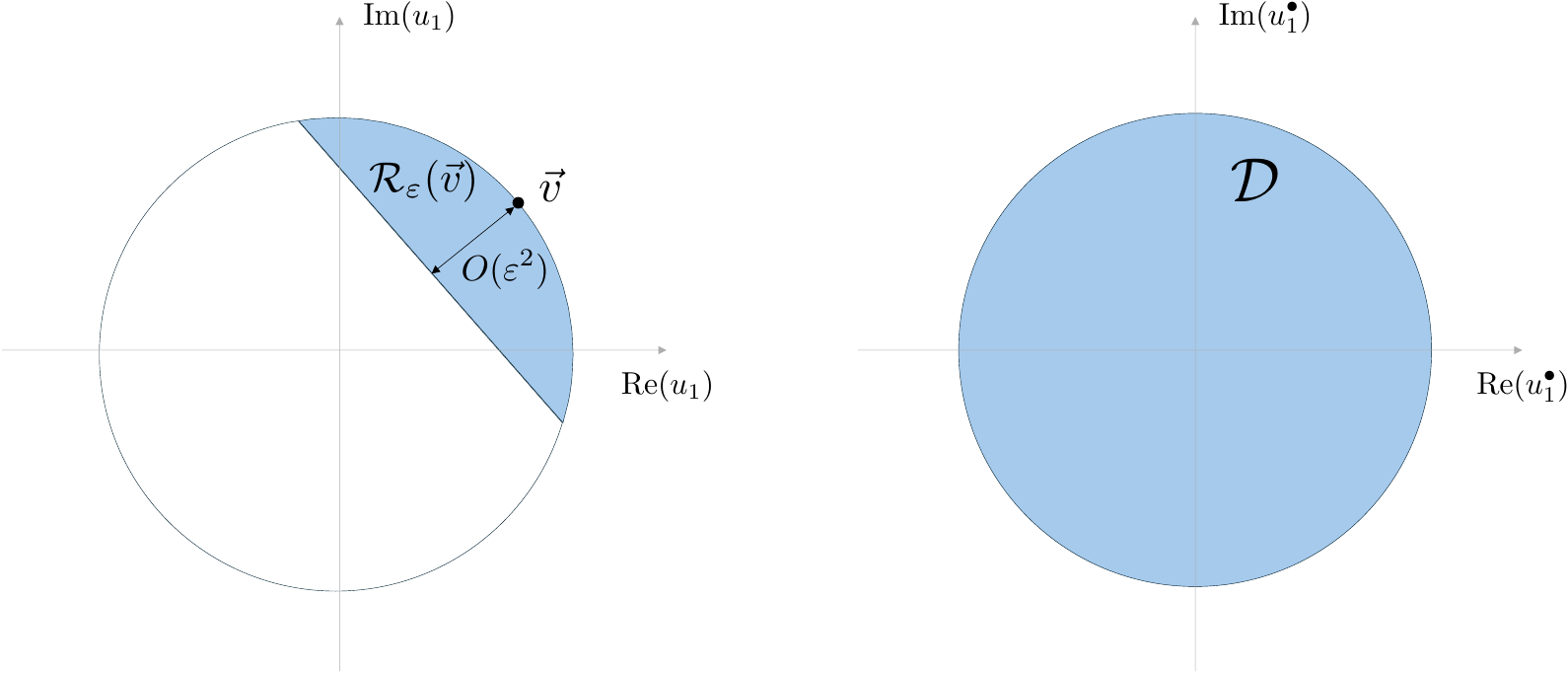}
    \caption{Illustration of $\mathcal{R}_\varepsilon(\vec{v})$ and $\mathcal{D}$ in $\R^2$. In practice, these sets are subsets of $\R^4$.}
    \label{fig: R_eps(v)}
\end{figure}

\begin{proof}
    Eq.\eqref{eq: u in A} follows directly from Conditions~\eqref{item: norm constraint} and \eqref{item: error constraint} of Problem~\ref{prob: enum u1 and u2 for fixed k}. 
    The complex conjugation map $(-)^\dag$ is an automorphism on $\Dzeta$ and acts as
    \begin{equation}
        \bigl(a + b\zeta + c\zeta^2 + d\zeta^3 \bigr)^\dag = a - d\zeta - c\zeta^2 - b\zeta^3.
    \end{equation}
    It is easy to verify that $(x^\dag)^\bullet = (x^\bullet)^\dag$ for all $x \in \Dzeta$. 
    So we obtain
    \begin{equation}
        u_1^\dag u_1 + u_2^\dag u_2 = 1 \Longleftrightarrow (u_1^\bullet)^\dag(u_1^\bullet) + (u_2^\bullet)^\dag(u_2^\bullet) = 1,
    \end{equation}
    which in turn implies Eq.~\eqref{eq: u^dot in B}.
\end{proof}

Instead of requiring conditions (\ref{item: norm constraint}) and (\ref{item: error constraint}) in Problem~\ref{prob: enum u1 and u2 for fixed k}, we enumerate $(u_1, u_2)^\text{t} \in \Dzeta^2$ satisfying $\vec{u} \in \mathcal{R}_\varepsilon(\vec{v})$ and $\vec{u}^\bullet \in \mathcal{D}$. 
We then search those candidates to obtain actual solutions to Problem~\ref{prob: enum u1 and u2 for fixed k}, i.e., those $u$ that satisfy conditions (\ref{item: norm constraint}) and (\ref{item: error constraint}).
From condition (\ref{item: lde constraint}) in Problem~\ref{prob: enum u1 and u2 for fixed k}, it follows that any candidate $(u_1, u_2)^\text{t}$ can written in the form
\begin{align}
\label{eq: u1 and u2 integer form}
    u_1 = \frac{a_1 + b_1\zeta + c_1\zeta^2 + d_1\zeta^3}{\sqrt{2}^k} \quad \text{and} \quad u_2 = \frac{a_2 + b_2\zeta + c_2\zeta^2 + d_2\zeta^3}{\sqrt{2}^k}
\end{align}
for some integers $a_1, b_1, c_1, d_1, a_2, b_2, c_2, d_2 \in \Z$.
Hence, these integers satisfy the following relation:
\begin{equation}
\label{eq: u u^dot in R_eps x D}
    \begin{pmatrix}
        \Vec{u} \\ \Vec{u}^\bullet
    \end{pmatrix} = 
    \begin{pmatrix}
        \Re(u_1) \\
        \Im(u_1) \\
        \Re(u_2) \\
        \Im(u_2) \\
        \Re(u_1^\bullet) \\
        \Im(u_1^\bullet) \\
        \Re(u_2^\bullet) \\
        \Im(u_2^\bullet)
    \end{pmatrix} = 
    \frac{1}{\sqrt{2}^k} \Sigma \begin{pmatrix}
        a_1 \\ b_1 \\ c_1 \\ d_1 \\
        a_2 \\ b_2 \\ c_2 \\ d_2
    \end{pmatrix} \in \mathcal{R}_\varepsilon(\vec{v}) \times \mathcal{D},
\end{equation}
where the matrix $\Sigma \in \R^{8 \times 8}$ is defined by
\begin{equation}
        \Sigma \coloneqq \left(\begin{array}{cccccccc}
        1 & \frac{1}{\sqrt{2}} & 0 & -\frac{1}{\sqrt{2}} & 0 & 0 & 0 & 0  \\
        0 & \frac{1}{\sqrt{2}} & 1 & \frac{1}{\sqrt{2}} & 0 & 0 & 0 & 0 \\
        0 & 0 & 0 & 0 & 1 & \frac{1}{\sqrt{2}} & 0 & -\frac{1}{\sqrt{2}} \\
        0 & 0 & 0 & 0 & 0 & \frac{1}{\sqrt{2}} & 1 & \frac{1}{\sqrt{2}} \\ 
        1 & -\frac{1}{\sqrt{2}} & 0 & \frac{1}{\sqrt{2}} & 0 & 0 & 0 & 0  \\
        0 & -\frac{1}{\sqrt{2}} & 1 & -\frac{1}{\sqrt{2}} & 0 & 0 & 0 & 0 \\
        0 & 0 & 0 & 0 & 1 & -\frac{1}{\sqrt{2}} & 0 & \frac{1}{\sqrt{2}} \\
        0 & 0 & 0 & 0 & 0 & -\frac{1}{\sqrt{2}} & 1 & -\frac{1}{\sqrt{2}} \\
    \end{array} \right).
\end{equation}
Note that $\Sigma$ is invertible, we can reduce Problem~\ref{prob: enum u1 and u2 for fixed k} to enumerating the integer points in the convex body 
\begin{equation}
\label{eq: dfn S eps k v}
    S_{\varepsilon, k}(\vec{v}) \coloneqq \sqrt{2}^k \Sigma^{-1} \bigl(\mathcal{R}_\varepsilon(\vec{v}) \times \mathcal{D} \bigr).
\end{equation}
Applying the algorithm from Problem~\ref{prop: integer enum algorithm} allows us to enumerate this set efficiently. 
We summarize the algorithm developed thus far for solving Problem~\ref{prob: sub dete synth} and analyze the computational cost of the algorithm. 
\begin{algo}
\label{algo: enum lattice sub dete synth}
    Given a single-qubit unitary 
    $V = \left( \begin{array}{rr}
        v_1 & -v_2^\dag \\
        v_2 &  v_1^\dag
    \end{array} \right)$
    , an approximation error $\varepsilon \in (0,1]$ and a $T$-count $t \in \N$, perform the following steps:
    \begin{enumerate}
        \item Set
        \begin{itemize}
            \item if $t$ is even, $k = \frac{t+2}{2}$ and $\vec{v} = \begin{pmatrix} v_1 \\ v_2 \end{pmatrix}$, 
            \item if $t$ is odd, $k = \frac{t+3}{2}$ and $\vec{v} = \sqrt{\zeta}\begin{pmatrix} v_1 \\ v_2 \end{pmatrix}$.
        \end{itemize}
        
        \item Enumerate the integer points in $S_{\varepsilon,k}(\vec{v}) \subseteq \R^8$ defined by Eq.~\eqref{eq: dfn S eps k v}, using the algorithm of Proposition~\ref{prop: integer enum algorithm}. 
        
        \item Using the results from Step~(b), construct candidate solutions to Problem~\ref{prob: enum u1 and u2 for fixed k} based on Eq~\eqref{eq: u1 and u2 integer form}, and then extract the actual solutions $\vec{u} = \begin{pmatrix} u_1 \\ u_2 \end{pmatrix}$ from among them.  
        
        \item For each $(u_1, u_2)^\text{t}$, define the single-qubit unitary
        \begin{equation}
            U = \begin{cases}
                \left( \begin{array}{rr}
                    u_1 & -u_2^\dag \\
                    u_2 &  u_1^\dag
                \end{array} \right) & \text{if $t$ is even,} \\[15pt]

                \left( \begin{array}{rr}
                    u_1 & -u_2^\dag \zeta \\
                    u_2 &  u_1^\dag \zeta
                \end{array} \right) & \text{if $t$ is odd,}
            \end{cases}
        \end{equation}
        and output those whose $T$-count is exactly $t$.
    \end{enumerate}
\end{algo}

\begin{prop}[Correctness and Termination of Algorithm~\ref{algo: enum lattice sub dete synth}]
\label{prop: correctness enum lattice sub dete synth}
    Algorithm~\ref{algo: enum lattice sub dete synth} necessarily outputs $\CliffordTOne_{\Tcount = t} \cap \mathbf{B}_\varepsilon(V)$ in finite time. 
\end{prop}
\begin{proof}
    By construction.     
\end{proof}

\begin{prop}[Complexity of Algorithm~\ref{algo: enum lattice sub dete synth}]
\label{prop: complexity enum lattice sub dete synth}
    Let $\CostAlg{\ref*{algo: enum lattice sub dete synth}}{\varepsilon, t}(V)$ denote the arithmetic operation count of Algorithm~\ref{algo: enum lattice sub dete synth}.  
    It holds that
    \begin{equation}
        \Expect[\CostAlg{\ref*{algo: enum lattice sub dete synth}}{\varepsilon, t}] = O\Bigl( (2^{2t}\varepsilon^5 + 2^t\varepsilon^3 + 1)\poly(t) \Bigr) \quad \text{as $1/\varepsilon,\ t \rightarrow \infty$}.
    \end{equation}
\end{prop}

\begin{proof}
    It is clear that the main computational cost of Algorithm~\ref{algo: enum lattice sub dete synth} is incurred by the enumeration of integer points in Step~(b). 
    Moreover, by Proposition~\ref{prop: integer enum algorithm}, for all $V \in \UC(2)$ we have
    \begin{equation}
        \CostAlg{\ref*{algo: enum lattice sub dete synth}}{\varepsilon, t}(V) = O\left(|S_{\varepsilon,k}(\vec{v}) \cap \Z^8| + \poly\log \bigl(\vol(S_{\varepsilon,k}(\vec{v})) \bigr) \right),
    \end{equation}
    where $\vec{v} \in \widetilde{S}^3$ and $k \in \N$ are defined by Step~(a). 
    The second term on the right-hand side is deterministically analyze, i.e., independent of $V$ and a straightforward volume calculation shows that $\vol(S_{\varepsilon,k}(\vec{v})) = O(\varepsilon^5 2^{4k})$ for any $V \in \UC(2)$. 
    Next, we analyze the first term. 
    From the bi-invariance of the Haar measure, we may treat $\vec{v}$ as a sample from the Haar measure $\mu_{\widetilde{S}^3}$ regardless of whether $t$ is even or odd. 
    Therefore, we have the following bound:
    \begin{equation}
    \label{eq: bound of expected value of runtime 1}
        \mathbb{E} \bigl[\CostAlg{\ref*{algo: enum lattice sub dete synth}}{\varepsilon, t}\bigr] = O\left( \mathbb{E}_{ \mu_{\widetilde{S}^3} }[N_{\varepsilon,k}] \right) + \poly(t),
    \end{equation}
    where the random variable $N_{\varepsilon, k}(\vec{v}) = |S_{\varepsilon, k}(\vec{v}) \cap \Z^8|$. 
    To analyze expectation value of $N_{\varepsilon,k}$, we define two finite subsets of $\C^2$:
    \newcommand{\Pepsk}{P_{\varepsilon,k}}
    \newcommand{\hatPepsk}{\widehat{P_{\varepsilon,k}}}
    \begin{align}
        \Pepsk &\coloneqq \Bigl\{ x \in \Zzeta^2 \ : \ (1-\varepsilon^2)2^k \leq \|x\|^2 \leq 2^k, \ \|x^\bullet\|^2 \leq 2^k \Bigr\} \quad \text{and} \\
        \hatPepsk &\coloneqq \Bigl\{ \tfrac{p}{\sqrt{2}^k} \ : \ p \in P_{\varepsilon,k}  \Bigr\}.
    \end{align} 
    If we identify $\C^2$ with $\R^4$, it is easy to see that that $N_{\varepsilon,k}(\vec{v}) = |\hatPepsk \cap \mathcal{R}_\varepsilon(\vec{v})|$ for any $\vec{v} \in \widetilde{S}^3$ from Eq.~\eqref{eq: u u^dot in R_eps x D}. 
    Moreover, if we define a set $B_\varepsilon(p) \coloneqq \left\{ \vec{v} \in S^3 \ | \ p \in \mathcal{R}_\varepsilon(\vec{v}) \right\}$ for $p \in \R^4$, then it is clear that 
    \begin{equation}
        N_{\varepsilon,k}(\vec{v}) = |\hatPepsk \cap \mathcal{R}_\varepsilon(\vec{v})| = \sum_{p \in \hatPepsk} 1_{B_\varepsilon(p)}(\vec{v}) \quad \text{for any $\vec{v} \in S^3$},
    \end{equation}
    where $1_A$ is the indicator function of $A$. 
    For any $p \in \hatPepsk$ and any $\vec{v} \in \widetilde{S}^3$, since $\|p\| \leq 1$ by definition of $\Pepsk$ and $\hatPepsk$, we have $|p \cdot \vec{v}| \leq \left| \frac{p}{\|p\|} \cdot \vec{v} \right|$. 
    This inequality implies that $B_\varepsilon(p) \subseteq B_\varepsilon\Bigl(\frac{p}{\|p\|} \Bigr) = \mathbf{B}_\varepsilon\left(\frac{p}{\|p\|} \right)$ for any $p \in \hatPepsk$, where $\Ball{r}{x}$ is the ball centered at $x \in \widetilde{S}^3$ and radius $r > 0$, induced by the metric $d_\diamond$. 
    Hence, we have
    \begin{align}
        \mathbb{E}_{\mu_{\widetilde{S}^3}}[N_{\varepsilon,k}] &= \sum_{p \in \hatPepsk} \mathbb{E}_{\mu_{\widetilde{S}^3}} \bigl[1_{B_\varepsilon(p)}\bigr] \\
        &= \sum_{p \in \hatPepsk} \Pr_{\vec{v} \sim \mu_{\widetilde{S}^3}} [\vec{v} \in B_\varepsilon(p)] \\
        &\leq \sum_{p \in \hatPepsk} \Pr_{\vec{v} \sim \mu_{\widetilde{S}^3}}\left[ \vec{v} \in \mathbf{B}_\varepsilon\left(\frac{p}{\|p\|}\right) \right] \\
        &= |\hatPepsk| \times \Pr_{\vec{v} \sim \mu_{\widetilde{S}^3}}\bigl[\vec{v} \in \Ball{\varepsilon}{e} \bigr] \\
        &= O\left( |\hatPepsk| \varepsilon^3 \right),
    \end{align}
    where $e$ is the identity element of $\widetilde{S}^3$. 

    We next analyze $|\hatPepsk|$ (in fact $|\Pepsk|$). To this end, we introduce two definitions:
    \begin{itemize}
        \item $r(m) \coloneqq \#\{ x \in \Zzeta^2 \ | \ \|x\|^2 = m \} \quad \text{for $m \in \Zsqrttwo$} \quad$ and,
        \item $I(x_0, x_1, y_0, y_1) \coloneqq \{ x \in \Zsqrttwo \ | \ x \in [x_0, x_1], \  x^\bullet \in [y_0, y_1] \} \quad  \text{for $x_0, x_1, y_0,y_1 \in \R$}$.
    \end{itemize}
    With these definitions in place, the cardinality of $\Pepsk$ can be written as follows:
    \begin{equation}
        |\Pepsk| = \sum_{m \in I \bigl((1-\varepsilon^2)2^k, \ 2^k, \ 0, \ 2^k\bigr)} r(m). 
    \end{equation}
    In Appendix~\ref{appendix: analysis of r(m)}, we prove $r(m) = O\bigl(|m m^\bullet| (\log \log|mm^\bullet| )^2\bigr)$ and Ross and Selinger~\cite{ross2016optimal} has shown $\#I \bigl((1-\varepsilon^2)2^k, \ 2^k, \ 0, \ 2^k\bigr) = O(2^{2k} \varepsilon^2 + 1)$. 
    Although Appendix~\ref{appendix: analysis of r(m)} does not directly prove this claim, note that it holds by virtue of the relation between $\Zzeta$ and $\Zsqrttwo$. 
    Furthermore, since $|m m^\bullet| \leq |m||m^\bullet| \leq 2^{2k}$ for every $m \in I \bigl((1-\varepsilon^2)2^k, \ 2^k, \ 0, \ 2^k\bigr)$, we have
    \begin{equation}
        |\Pepsk| = O\bigl( (2^{4k} \varepsilon^2 + 2^{2k})\polylog (k)\bigr).
    \end{equation}
    Summarizing the preceding discussion, we can see that
    \begin{equation}
        \mathbb{E}_{\mu_{\widetilde{S}^3}}\bigl[N_{\varepsilon,k} \bigr] = O\left( (2^{4k}\varepsilon^5 + 2^{2k}\varepsilon^3)\polylog (k) \right).
    \end{equation}
    Recalling that $k = \tfrac{t}{2} + O(1)$ and combining this equation with the Eq.~\eqref{eq: bound of expected value of runtime 1} proves the claim.
\end{proof}

\subsection{The algorithm based on divide-and-conquer \label{subsec: the algorithm based on divide-and-conquer}}

We will now propose a new algorithm for solving Problem~\ref{prob: sub dete synth} that combines a divide-and-conquer framework with the algorithm introduced Section~\ref{subsec: reduction to integer point enum}.
In preparation for showing that Problem~\ref{prob: sub dete synth} can be divided into smaller subproblems, we first introduce the Matsumoto-Amano normal form \cite{giles2013remarks, matsumoto2008representation}.
\begin{prop}[Matsumoto-Amano normal form]
\label{prop: matsumoto-amano normal form}
    For any $U \in \CliffordTOne$, there exist a unique factorization
    \begin{equation}
    \label{eq: matsumoto-amano factorization}
        U = T^{a_0} \left( \prod_{i=1}^m HS^{b_i}T \right) C,
    \end{equation}
    where $m \in \N$, $a_0, b_1, \dots, b_m \in \{0,1\}$ and $C \in \mathcal{C}_1$. 
    Moreover, this representation uses a minimum number of $T$ gates, namely
    \begin{equation}
        \Tcount(U) = a_0 + m.
    \end{equation}
\end{prop}

The following lemma shows that a solution to Problem~\ref{prob: sub dete synth} can be constructed from some solutions to Problem~\ref{prob: sub dete synth} with a smaller instance. 
\begin{lem}
\label{lem: divide S(V,eps,t)}
    For $n \in \N$, define the subset of single-qubit Clifford+$T$ gates
    \begin{equation}
    \label{eq: define L(n)}
        L_n \coloneqq \begin{cases}
            \{ I \} & \text{$\mathrm{if}$ $n = 0$} \\[6pt]
            \displaystyle
            \Bigl\{ \prod_{i=1}^{n} HS^{b_i}T \ \Bigm| \ b_i \in \{0,1\} \Bigr\}
            \ \bigcup \
            \Bigl\{ T \prod_{i=1}^{n-1} HS^{b_i}T \ \Bigm| \ b_i \in \{0,1\} \Bigr\} & \text{$\mathrm{if}$ $n \geq 1$}
        \end{cases}.
    \end{equation}
    Then, for any $\varepsilon > 0$, any natural numbers $t, t'$ with $t' \leq t$, we have
    \begin{equation}
    \label{eq: divide S(V,eps,t)}
        \CliffordTOne_{\Tcount = t} \cap \Ball{\varepsilon}{V} \subseteq \bigcup_{U_L \in L_{t'}} U_L \left( \CliffordTOne_{\Tcount = t-t'} \cap \mathbf{B}_\varepsilon(U_L^\dag V) \right).
    \end{equation}
\end{lem}

\begin{proof}
    For $n \in \N$, define subsets of single-qubit Clifford+$T$ gates
    \begin{align}
        R_n &\coloneqq \Bigl\{ \Bigl(\prod_{i=1}^{n} HS^{b_i}T \Bigr) C \ \Bigm| \ b_i \in \{0,1\}, \ C \in \mathcal{C}_1 \Bigr\}. 
    \end{align}
    From Proposition~\ref{prop: matsumoto-amano normal form}, for every natural numbers $t$ and $t' \leq t$, we can see that
    \begin{equation}
        \CliffordTOne_{\Tcount = t} = \Bigl\{ U_L U_R \ \Bigm| \ U_L \in L_{t'}, U_R \in R_{t-t'} \Bigr\}.
    \end{equation}
    Similarly, from Proposition~\ref{prop: matsumoto-amano normal form} $R_{t-t'} \subseteq \CliffordTOne_{\Tcount = t-t'}$. 
    Therefore, we obtain
    \begin{align}
        \CliffordTOne_{\Tcount = t} \cap \Ball{\varepsilon}{V} 
        &=
        \Bigl\{ U \in \CliffordTOne_{\Tcount = t} \ \Bigm| \ \diamonddist{\mathcal{U}}{\mathcal{V}} < \varepsilon \Bigr\} \\
        &= \Bigl\{ U_L U_R \Bigm| U_L \in L_{t'}, \ U_R \in R_{t-t'}, \ \diamonddist{\mathcal{U}_L \mathcal{U}_R}{\mathcal{V}} < \varepsilon \Bigr\} \\
        &= \bigcup_{U_L \in L_{t'}} U_L \Bigl\{ U_R \in R_{t-t'} \ \Bigm| \diamonddist{\mathcal{U}_R}{\mathcal{U}_L^\dag\mathcal{V}} < \varepsilon  \Bigr\} \\
        &\subseteq \bigcup_{U_L \in L_{t'}} U_L \Bigl\{ U_R \in \CliffordTOne_{\Tcount = t-t'} \ \Bigm| \diamonddist{\mathcal{U}_R}{\mathcal{U}_L^\dag\mathcal{V}} < \varepsilon  \Bigr\} \\
        &= \bigcup_{U_L \in L_{t'}} U_L \left( \CliffordTOne_{\Tcount = t - t'} \cap \mathbf{B}_\varepsilon(U_L^\dag V) \right).
    \end{align}
\end{proof}

Using Eq.~\eqref{eq: divide S(V,eps,t)}, we design a divide-and-conquer algorithm to solve Problem~\ref{prob: sub dete synth}.
Concretely, for each $U_L \in L_{t'}$ we compute $\CliffordTOne_{t-t'} \cap \mathbf{B}_\varepsilon(U_L^\dag V)$ with Algorithm~\ref{algo: enum lattice sub dete synth} and then merge all such results. 
Next, we discuss how to choose $t'$. 
According to Proposition~\ref{prop: complexity enum lattice sub dete synth}, each call to Algorithm~\ref{algo: enum lattice sub dete synth} requires $O\left( (2^{2(t-t')}\varepsilon^5 + 2^{t-t'}\varepsilon^3 + 1)\poly(t) \right)$ arithmetic operations in expectation. 
Intuitively, since $|L_{t'}| = O(2^{t'})$ by the definition, the divide-and-conquer algorithm based on Eq.~\eqref{eq: divide S(V,eps,t)} requires $O\left( (2^{2t-t'}\varepsilon^5 + 2^t\varepsilon^3 + 2^{t'})\poly(t) \right)$ arithmetic operations in expectation. 
Therefore, by applying the AM-GM inequality, we see that the optimal choice is
\begin{equation}
    t' = t - \frac52 \log_2 (1/\varepsilon). 
\end{equation}
We describe the algorithm in detail and analyze its computational cost.

\begin{algo}
\label{algo: divide and conquer sub dete synth}
    Given a single-qubit unitary $V$, an approximation error $\varepsilon \in (0,1]$ and $t \in \N$, perform the following steps:
    \begin{enumerate}
        \item Set $t' = \max\big( 0, \  \lceil t - 5/2\log_2 (1/\varepsilon) \rfloor \big)$.
        \item For each $U_L \in L_{t'}$, run the Algorithm~\ref{algo: enum lattice sub dete synth} with input $\bigl(U_L^{-1} V, \ \varepsilon, \ t-t'\bigr)$.
        \item Merge the results from Step~(b) using Eq.~\eqref{eq: divide S(V,eps,t)} and output those whose $T$-count is exactly $t$.  
    \end{enumerate}
\end{algo}

\begin{prop}[Correctness and Termination of Algorithm~\ref{algo: divide and conquer sub dete synth}]
\label{prop: correctness divide and conquer sub dete synth}
    Algorithm~\ref{algo: divide and conquer sub dete synth} necessarily outputs $\CliffordTOne_{\Tcount = t} \cap \mathbf{B}_\varepsilon(V)$ in finite time. 
\end{prop}
\begin{proof}
    This follows immediately from Proposition~\ref{prop: correctness enum lattice sub dete synth} and Lemma~\ref{lem: divide S(V,eps,t)}.    
\end{proof}

\begin{prop}[Complexity of Algorithm~\ref{algo: divide and conquer sub dete synth}]
\label{prop: complexity divide and conquer sub dete synth}
    Let $\CostAlg{\ref*{algo: divide and conquer sub dete synth}}{\varepsilon, t}(V)$ denote the arithmetic operation count of Algorithm~\ref{algo: divide and conquer sub dete synth}. 
    It holds that
    \begin{equation}
        \Expect[\CostAlg{\ref*{algo: divide and conquer sub dete synth}}{\varepsilon, t}] = 
        \begin{cases}
            O\left( ( 2^{2t}\varepsilon^5 + 2^{t}\varepsilon^3 + 1) \poly(t) \right) & \mathrm{if} \ \ t < \frac{5}{2}\log_2(1/\varepsilon) \\[2mm]
            O\left( 2^{t}\varepsilon^{5/2} \poly(t) \right) & \mathrm{if} \ \ t \geq \frac{5}{2}\log_2(1/\varepsilon) 
        \end{cases} \quad \text{as $1/\varepsilon,\; t \rightarrow \infty$}.
    \end{equation}
\end{prop}

\begin{proof}
    It is clear that
    \begin{equation}
        \CostAlg{\ref*{algo: divide and conquer sub dete synth}}{\varepsilon,t}(V) = O\left(\sum_{U_L \in L_{t'}} \CostAlg{\ref*{algo: enum lattice sub dete synth}}{\varepsilon,t-t'}\bigl(U_L^{-1}V \bigr) \right) \quad \text{as $1/\varepsilon,\; t \rightarrow \infty$}
    \end{equation}
    for all $V \in \UC(2)$. 
    Therefore, by applying the bi-invariance of the Haar measure and Proposition~\ref{prop: complexity enum lattice sub dete synth}, we have the following bound:
    \begin{align}
        \mathbb{E} \bigl[ \CostAlg{\ref*{algo: divide and conquer sub dete synth}}{\varepsilon,t} \bigr] &= O\left( |L_{t'}| \times  \mathbb{E} \bigl[ \CostAlg{\ref*{algo: enum lattice sub dete synth}}{\varepsilon,t-t'} \bigr]   \right) \\
        &= O\left( (2^{2t-t'}\varepsilon^5 + 2^{t}\varepsilon^3 + 2^{t'})\poly(t) \right) \quad \text{as $1/\varepsilon,\; t \rightarrow \infty$}.
    \end{align}
    By the definition of $t'$ in Step~(a), we see that the claim holds. 
\end{proof}

\subsection{Overall algorithm}
Finally, we construct an algorithm that solves the $T$-optimal deterministic Clifford+$T$ synthesis for single-qubit unitaries. 
\begin{algo}[$T$-optimal Deterministic Synthesis]
\label{algo: dete synth}
    Given a single-qubit unitary $V = \left( \begin{array}{rr}
        v_1 & -v_2^\dag \\
        v_2 &  v_1
    \end{array} \right)$ and an approximation error $\varepsilon > 0$, perform the following steps:
    \begin{enumerate}
        \item If $\varepsilon > 1$, output $I_2$ and terminate the algorithm. 
        \item Set $t \leftarrow 0$. 
        \item Compute $\CliffordTOne_{\Tcount = t} \cap \mathbf{B}_\varepsilon(V)$ using Algorithm~\ref{algo: divide and conquer sub dete synth}. 
        \item If the result of Step~(b) is empty, update $t \leftarrow t+1$ and repeat Step~(b); otherwise, output an arbitrary element of $\CliffordTOne_{\Tcount = t} \cap \mathbf{B}_\varepsilon(V)$.
    \end{enumerate}
\end{algo}

\begin{prop}[Correctness and Termination of Algorithm~\ref{algo: dete synth}]
\label{prop: correctness dete synth}
    Algorithm~\ref{algo: dete synth} necessarily outputs a solution to Problem~\ref{prob: dete synth} in finite time. 
\end{prop}
\begin{proof}
    The optimality with respect to the $T$-count is evident. Moreover, the correctness and termination of Step~(b) follows from Proposition~\ref{prop: correctness divide and conquer sub dete synth}. 
\end{proof}

Next, we analyze the runtime and $T$-count of Algorithm~\ref{algo: dete synth}. 
To derive a theoretical bound, we must first establish an upper bound on the variable $t$ in Algorithm~\ref{algo: dete synth}. 
This discussion addresses the fundamentally important question of how many $T$ gates are required to achieve an $\varepsilon$-approximation of a single-qubit unitary. 
Parzanchevski and Sarnak~\cite{parzanchevski2018super} proved the following proposition concerning this problem. 
\begin{lem}[Proposition~3.1 of \cite{parzanchevski2018super}]
\label{lem: Clifford+T covering bound}
    It holds that
    \begin{equation}
        \Pr \left[ V \notin \bigcup_{U \in \CliffordTOne_{\Tcount \leq t}} \mathbf{B}_\varepsilon(U)  \right]
           = O\left( \frac{t^2}{2^t \varepsilon^3} \right) \quad \text{as $1/\varepsilon,\; t \rightarrow \infty$}.
    \end{equation}
\end{lem}

\begin{thm}[Complexity of Algorithm~\ref{algo: dete synth}]
\label{thm: complexity dete synth}
    Let $\TimeAlg{\ref*{algo: dete synth}}{\varepsilon}(V)$ be the runtime of Algorithm~\ref{algo: dete synth}. 
    For any $\gamma > 0$, it holds that
    \begin{equation}
        \TimeAlg{\ref*{algo: dete synth}}{\varepsilon} = \Owhp(\varepsilon^{-1/2 - \gamma}) \quad \text{as $1/\varepsilon \rightarrow \infty$}.
    \end{equation}
\end{thm}

\begin{proof}
    Let $\Tcount_\varepsilon(V)$ is the maximum value of $t$ in Algorithm~\ref{algo: dete synth}. 
    It is easy to see that
    \begin{equation}
        \TimeAlg{\ref*{algo: dete synth}}{\varepsilon} = O\left(  \sum_{t=0}^{\Tcount_\varepsilon} \CostAlg{\ref*{algo: divide and conquer sub dete synth}}{\varepsilon,t} \right) \times \polylog(1/\varepsilon) \quad \text{as $1/\varepsilon \rightarrow \infty$}.
    \end{equation} 
    Note that each arithmetic operation is performed with $O(\log(1/\varepsilon))$-bit precision. 
    By Proposition~\ref{prop: correctness dete synth} and Lemma~\ref{lem: Clifford+T covering bound}, we have
    \begin{equation}
    \label{eq: probability of Tcount_eps > alpha}
        \Pr\bigl[ \Tcount_\varepsilon(V) > \alpha \bigr] = \Pr\left[ V \notin \bigcup_{U \in \CliffordTOne_{\Tcount \leq \alpha}} \mathbf{B}_\varepsilon(U) \right] = O\left( \frac{\alpha^2}{2^\alpha \varepsilon^3} \right) \quad \text{as $1/\varepsilon,\; \alpha \rightarrow \infty$}.
    \end{equation}
    For any $\gamma_1 > 0$, it holds that
    \begin{align}
        \Pr\left[ \sum_{t=0}^{\Tcount_\varepsilon(V)} \CostAlg{\ref*{algo: divide and conquer sub dete synth}}{\varepsilon,t}(V) > \sum_{t=0}^{(3+\gamma_1)\log_2(1/\varepsilon)} \CostAlg{\ref*{algo: divide and conquer sub dete synth}}{\varepsilon, t}(V) \right] 
        &= 
        \Pr[\Tcount_\varepsilon(V) > (3+\gamma_1)\log_2(1/\varepsilon)] \\
        &= O\bigl(\varepsilon^{\gamma_1} \polylog(1/\varepsilon)\bigr).
    \end{align}
    Therefore, we obtain
    \begin{equation}
        \sum_{t=0}^{\Tcount_\varepsilon} \CostAlg{\ref*{algo: divide and conquer sub dete synth}}{\varepsilon,t} = \Owhp\left( \sum_{t=0}^{(3+\gamma_1)\log_2(1/\varepsilon)} \CostAlg{\ref*{algo: divide and conquer sub dete synth}}{\varepsilon, t} \right) \quad \text{as $1/\varepsilon \rightarrow \infty$}. 
    \end{equation}
    By applying Markov's inequality and Proposition~\ref{prop: complexity divide and conquer sub dete synth}, for any $\gamma_2 > 0$, we have
    \begin{align}
        \Pr[\sum_{t=0}^{(3+\gamma_1)\log_2(1/\varepsilon)} \CostAlg{\ref*{algo: divide and conquer sub dete synth}}{\varepsilon, t}(V) > \varepsilon^{-1/2 - \gamma_1 - \gamma_2}] &\leq \varepsilon^{1/2 + \gamma_1 + \gamma_2} \sum_{t=0}^{(3+\gamma_1)\log_2(1/\varepsilon)}\Expect[\CostAlg{\ref*{algo: divide and conquer sub dete synth}}{\varepsilon, t}] \\
        &= \varepsilon^{1/2 + \gamma_1 + \gamma_2} \times O\left( \varepsilon^{-1/2 - \gamma_1}\polylog(1/\varepsilon) \right) \\
        &= O\left( \varepsilon^{\gamma_2}\polylog(1/\varepsilon) \right). 
    \end{align}
    It implies that
    \begin{equation}
        \sum_{t=0}^{(3+\gamma_1)\log_2(1/\varepsilon)} \CostAlg{\ref*{algo: sub prob synth}}{\varepsilon,t} = \Owhp(\varepsilon^{-1/2 - \gamma_1 - \gamma_2}). 
    \end{equation}
    From the transitivity of $\Owhp$, we complete the proof.
\end{proof}

\begin{thm}[Bound on the $T$-count for Deterministic Synthesis]
\label{thm: upper bound on T-count for Deterministic}
    Let $\Tcount_\varepsilon(V)$ be the $T$-count of a solution to Problem~\ref{prob: dete synth}. 
    It holds that 
    \begin{equation}
        \Tcount_\varepsilon = 3\log_2(1/\varepsilon) + \owhp(\log_2(1/\varepsilon)) \quad \text{as $1/\varepsilon \rightarrow \infty$}.
    \end{equation}
\end{thm}
\begin{proof}
    By Eq.~\eqref{eq: probability of Tcount_eps > alpha}, for any $\gamma > 0$, we have
    \begin{equation}
        \Pr[\Tcount_\varepsilon(V) > (3+\gamma)\log_2(1/\varepsilon)] = O\left( \varepsilon^{\gamma}\polylog(1/\varepsilon) \right).  
    \end{equation}
    Next, we prove the lower bound. 
    For any $\gamma > 0$, we have
    \begin{align}
        \Pr\left[\Tcount_\varepsilon(V) \leq (3 - \gamma)\log_2(1/\varepsilon)\right] 
        &= 
        \Pr[V \in \bigcup_{U \in \CliffordTOne_{\Tcount \leq (3-\gamma)\log_2(1/\varepsilon)}} \Ball{\varepsilon}{U} ] \\ 
        &\leq 
        \sum_{U \in \CliffordTOne_{\Tcount \leq (3-\gamma)\log_2(1/\varepsilon)}} \Pr[V \in \Ball{\varepsilon}{U}] \\
        &= \left| \CliffordTOne_{\Tcount \leq (3-\gamma)\log_2(1/\varepsilon)} \right| \times \Ball{\varepsilon}{I_2} \\
        &= O(\varepsilon^{-3 + \gamma} \times \varepsilon^{3}) = O(\varepsilon^\gamma).
    \end{align}
    Note that, by Matsumoto-Amano normal form, we can see the cardinality of $\CliffordTOne_{\Tcount \leq n}$.
    Thus, the proof is complete.
\end{proof}

\section{Probabilistic Synthesis\label{sec: Probabilistic Synthesis}}

In this section, we consider the approximation of a target unitary by probabilistic mixtures of Clifford+$T$ gates, realized as CPTP maps that apply Clifford+$T$ gates sampled according to a specified probability distribution.
Several studies~\cite{kliuchnikov2023shorter, campbell2017shorter, akibue2024probabilistic} have indicated that performing probabilistic mixing relaxes the approximation accuracy required of each unitary component in the mixture. 

\begin{prob}[$T$-optimal Probabilistic Synthesis]
\label{prob: prob synth}
    Given a single-qubit unitary $V$ and an approximation error $\varepsilon > 0$, find a finite set of single-qubit Clifford+$T$ gates $\{ U_x \}_{x \in X}$ with minimum $\max_{x \in X} \Tcount(U_x)$ and a probability distribution $p$ over $X$ such that
    \begin{equation}
        \diamonddist{\sum_{x \in X} p(x) \mathcal{U}_x}{\mathcal{V}} < \varepsilon.
    \end{equation}
\end{prob}

Analogously to the $T$-optimal deterministic Clifford+$T$ synthesis algorithm in Section~\ref{sec: Deterministic Synthesis}, we instead focus on the following subproblem.

\begin{prob}[Fixed $T$-count Probabilistic Synthesis]
\label{prob: sub prob synth}
    Given a single-qubit unitary $\mathcal{V}$, a precision $\varepsilon > 0$ and $t \in \N$, find a finite set of single-qubit Clifford+$T$ gates $\{ U_x \}_{x \in X}$ and a probability distribution $p$ over $X$ such that $\max_{x \in X} \Tcount(U_x) = t$ and $\diamonddist{\sum_{x \in X} p(x) \mathcal{U}_x}{\mathcal{V}} < \varepsilon$. 
\end{prob}

\subsection{Reduction to the deterministic synthesis}
The following proposition provides a method for computing the optimal probability distribution when a set of Clifford+$T$ gates $\{U_x\}_{x \in X}$ to be mixed is given.
\begin{lem}[Proposition~3.1 of \cite{akibue2024probabilistic}]
\label{lem: compute optimal prob dist through SDP}
    Let $\mathcal{A} : \StateSpace(n) \rightarrow \StateSpace(m)$ be a CPTP mapping and $\bigl\{\mathcal{B}_x : \StateSpace(n) \rightarrow \StateSpace(m)\bigr\}_{x \in X}$ be a finite set of CPTP mappings. Then, the optimal distance $\min_p \diamonddist{\mathcal{A}}{\sum_{x \in X} p(x)\mathcal{B}_x}$ and the optimal probability distribution $p$, which minimizes the distance, can be computed with the following SDP:
    \begin{equation}
    \label{eq: SDP compute optimal distribution}
        \begin{tabular}{rlcrl}
        \multicolumn{2}{c}{\underline{{\rm Primal problem}}} &\ \ \ \ \ \ \ \ \ \ \ \ \ \  
         &\multicolumn{2}{c}{\underline{{\rm Dual problem}}}\\
        {\rm maximize:}&$\tr[J(\mathcal{A})T]-t$&&{\rm minimize:}&$r\in\mathbb{R}$ \\
        {\rm subject to:}& $0\leq T\leq\rho\otimes\mathbb{I}_{m}$,&&
        {\rm subject to:}& $S\geq0\wedge S\geq J\left(\mathcal{A}-\sum_{x\in X}p(x)\mathcal{B}_x\right)$,\\
        &$\rho\in\mathbb{I}_n$&&&$r\mathbb{I}_n\geq\tr_{\C^m}[S]$,\\
        &$\forall x\in X,\tr[J(\mathcal{B}_x)T]\leq t$.&&&$\forall x\in X,p(x)\geq0$,\\
        &&&&$\sum_{x\in X}p(x)\leq 1$.
        \end{tabular} 
    \end{equation}
\end{lem}
If $n$ and $m$ are constants, the SDP in Lemma~\ref{lem: compute optimal prob dist through SDP} admits an $\varepsilon$-precision solution in $\poly \bigl(|X|,\; \log (1/\varepsilon)\bigr)$ time~\cite{lovasz2003semidefinite}. 
Accordingly, applying Lemma~\ref{lem: compute optimal prob dist through SDP} with $\mathcal{A} = \mathcal{V}$ and $\{ \mathcal{B}_x \}_{x \in X} = \CliffordTOne_{\Tcount \leq t}$ yields a solution to Problem~\ref{prob: sub prob synth}. 
However, since $|\CliffordTOne_{\Tcount \leq t}| = O(2^t)$ by the Matsumoto-Amano normal, the brute-force method is impractical. 
From a convex-geometric perspective, a probabilistic mixture of single-qubit unitaries can be viewed as a convex combination in a Euclidean space of fixed dimension. 
By Crath\'{e}odory's theorem~\cite{Carathodory1911berDV}, this implies that the SDP of Lemma~\ref{lem: compute optimal prob dist through SDP} needs only involve a subset $\hat{X} \subseteq X$ of constant size. 
The following lemma shows that, for probabilistic mixtures of single-qubit unitaries, the SDP of Lemma~\ref{lem: compute optimal prob dist through SDP} can be optimized using only those unitaries lying in a neighborhood of the target unitary $V$. 
\begin{lem}
\label{lem: min_X equal min_hatX}
    Let $\{U_x \}_{x \in X}$ be a finite set of single-qubit unitaries, $V$ be a single-qubit unitary, and $\delta > 0$. 
    Define
    \begin{equation}
        \hat{X} \coloneqq \Bigl\{ x \in X \ \Bigm| \ \diamonddist{\mathcal{U}_x}{\mathcal{V}} < 2\delta \Bigr\},
    \end{equation}
    i.e., $\{ U_x \}_{x \in \hat{X}} = \{ U_x \}_{x \in X} \cap \Ball{2\delta}{V}$.
    If $\{U_x \}_{x \in \hat{X}}$ is a $\delta$-covering of $\Ball{\delta}{V}$, then it holds that
    \begin{equation}
    \label{eq: min_p over X equal min_p over Xhat leq delta^2}
        \min_{p \in \ProbDist(X)} \diamonddist{\sum_{x \in X} p(x) \mathcal{U}_x}{\mathcal{V}}
         = 
         \min_{\hat{p} \in \ProbDist(\hat{X})} \diamonddist{\sum_{x \in \hat{X}} \hat{p}(x) \mathcal{U}_x}{\mathcal{V}} 
         < 
         \delta^2.
    \end{equation}
\end{lem}
\begin{proof}
    By Eq.~(45) in the proof of Lemma~5.3 of \cite{akibue2024probabilistic}, it is shown that
   \begin{align}
        \label{eq: prob minimization on X primal = dual}
        \min_{p \in \ProbDist(X)} \diamonddist{\sum_{x \in X} p(x) \mathcal{U}_x}{\mathcal{V}}
        &= 
        \max_{W \in \UC(2)} \left(  \min_{x \in X} \diamonddist{\mathcal{W}}{\mathcal{U}_x}^2  - \diamonddist{\mathcal{W}}{\mathcal{V}}^2 \right), \ and \\
        \label{eq: prob minimization on Xhat primal = dual}
         \min_{\hat{p} \in \ProbDist(\hat{X})} \diamonddist{\sum_{x \in \hat{X}} \hat{p}(x) \mathcal{U}_x}{\mathcal{V}} 
         &= 
        \max_{W \in \UC(2)} \left(  \min_{x \in \hat{X}} \diamonddist{\mathcal{W}}{\mathcal{U}_x}^2  - \diamonddist{\mathcal{W}}{\mathcal{V}}^2 \right).
   \end{align}
   Moreover, when $\{ U_x \}_{x \in X}$ is a $\delta$-covering of $\Ball{\delta}{V}$, Eq.~(47) in the same proof also establishes that
   \begin{equation}
   \label{eq: W notin Ball => d(W,Ux) <= d(W,V)}
       \forall W \notin \mathbf{B}_\delta(V), \ \min_{x \in \hat{X}} \diamonddist{\mathcal{W}}{\mathcal{U}_x} \leq \diamonddist{\mathcal{W}}{\mathcal{V}}. 
   \end{equation}
   Let $W$ maximize the R.H.S. of Eq.~\eqref{eq: prob minimization on X primal = dual}. 
   Due to Eq.~\eqref{eq: W notin Ball => d(W,Ux) <= d(W,V)}, we can assume $W \in \Ball{\delta}{V}$. 
    Then, by the triangle inequality and the cover condition of $\{ U_x \}_{x \in \hat{X}}$, for all $x \in X \setminus \hat{X}$, we have
    \begin{equation}
        \diamonddist{\mathcal{W}}{\mathcal{U}_x} \geq \diamonddist{\mathcal{U}_x}{\mathcal{V}} - \diamonddist{\mathcal{W}}{\mathcal{V}} > 2\delta - \delta = \delta
    \end{equation}
    Therefore, due to the cover condition of $\{ U_x \}_{x \in \hat{X}}$, there exists $x_0 \in \hat{X}$ such that $\diamonddist{\mathcal{W}}{\mathcal{U}_{x_0}} = \min_{x \in X} \diamonddist{\mathcal{W}}{\mathcal{U}_x}$. 
    This implies Eq.~\eqref{eq: prob minimization on X primal = dual} $\geq$     Eq.~\eqref{eq: prob minimization on Xhat primal = dual}. 
    The converse inequality is trivial. 
    Moreover, since $\min_{x \in X} \diamonddist{\mathcal{W}}{\mathcal{U}_x} < \delta$, we obtain the inequality of Eq.~\eqref{eq: min_p over X equal min_p over Xhat leq delta^2}.
\end{proof}

\begin{figure}
    \centering
    \includegraphics[width=0.5\linewidth]{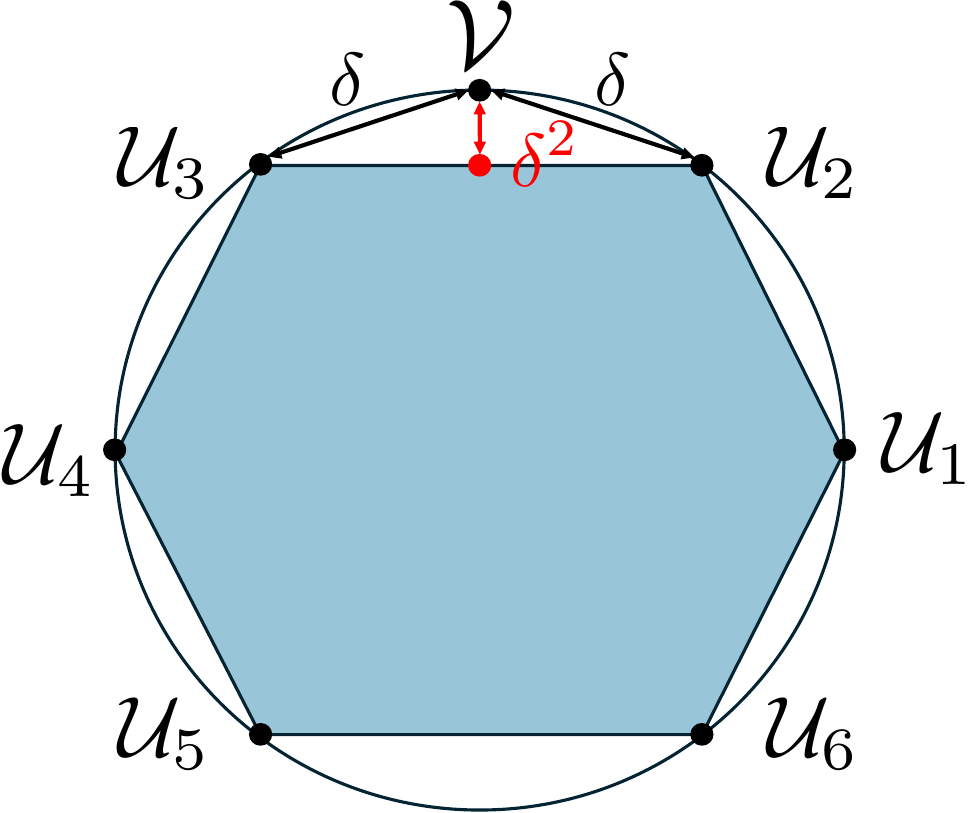}
    \caption{
    Illustration of Lemma~\ref{lem: min_X equal min_hatX} in $\R^2$ (in practice, a probabilistic mixture of single-qubit unitaries corresponds to a convex combination in $\R^{10}$). 
    The light blue region represents the set of CPTP mappings that can be realized as probabilistic mixtures of $\{U_x\}_{x \in \{1,2,3,4,5,6\} }$. 
    Among them, the CPTP mapping (red point) closest to the target unitary $V$ can be constructed by a probabilistic mixture of only $U_2$ and $U_3$ in the neighborhood of $V$. 
    Then, the optimal probabilistic mixture improves quadratically upon the individual approximation error $\delta$.
    }
    \label{fig:mix_unitary}
\end{figure}

Accordingly, setting $\{ U_x \}_{x \in X} = \CliffordTOne_{\Tcount \leq t}$, we consider an algorithm for Problem~\ref{prob: sub prob synth} that first finds a $\delta > 0$ satisfying the condition of Lemma~\ref{lem: min_X equal min_hatX} and then computes the optimization problem on the middle term of Eq.~\eqref{eq: min_p over X equal min_p over Xhat leq delta^2}. 
The following lemma states that there exists a polynomial-time algorithm that decides exactly whether the conditions of Lemma~\ref{lem: min_X equal min_hatX} hold.
\begin{lem}
\label{lem: covering verification algorithm}
    Let $\{U_x\}_{x \in \hat{X}}$ be a finite set of single-qubit unitaries, 
    $V$ be a single-qubit unitary, and $\delta \in (0, \frac12]$. 
    There exists an algorithm that, using $O\bigl( |\hat{X}|^4 \bigr)$ arithmetic operations, decides whether $\{ U_x \}_{x \in \hat{X}}$ is a $\delta$-covering of $\Ball{\delta}{V}$.
\end{lem}

\begin{proof}
    See Appendix~\ref{appendix: proof of covering verification algorithm}.
\end{proof}

Lemma~\ref{lem: covering verification algorithm} enables us to design an algorithm that iteratively increases $\delta$ until the conditions are satisfied. 
The details of initialization and update rule of $\delta$ are shown in Algorithm~\ref{algo: sub prob synth}.
\begin{algo}
\label{algo: sub prob synth}
    Given a single-qubit unitary $V$ and $t \in \N$, perform the following steps:
    \begin{enumerate}
        \item Set $\delta =  2^{-t/3 + c_1}$. 
        \item Compute $\CliffordTOne_{\Tcount \leq t} \cap \mathbf{B}_{2\delta}(V)$ using Algorithm~\ref{algo: divide and conquer sub dete synth}.
        \item If $2\delta > 1$, output $\CliffordTOne_{\Tcount \leq t} \cap \mathbf{B}_{2\delta}(V)$ and terminate the algorithm.
        \item Check whether $\CliffordTOne_{\Tcount \leq t} \cap \mathbf{B}_{2\delta}(V)$ is a $\delta$-covering of $\mathbf{B}_\delta(V)$ using the algorithm of Lemma~\ref{lem: covering verification algorithm}.
        
        \item If Step~(c) holds, outputs $\CliffordTOne_{\Tcount \leq t} \cap \mathbf{B}_{2\delta}(V)$; otherwise, update $\delta \leftarrow 2^{c_2} \delta$ and repeat Step~(b). 
    \end{enumerate}
    Here, $c_1$ and $c_2$ are positive constant parameters of this algorithm. 
\end{algo}

\begin{prop}[Correctness and Termination of Algorithm~\ref{algo: sub prob synth}]
\label{prop: correctness of sub prob synth}
    Let $\{U_x\}_{x \in X} = \CliffordTOne_{\Tcount \leq t}$ and let $\{U_x\}_{x \in \hat{X}}$ be the output of Algorithm~\ref{algo: sub prob synth}. 
    Then, it holds that
    \begin{equation}
        \min_{p \in \ProbDist(X)} \diamonddist{\sum_{x \in X} p(x) \mathcal{U}_x}{\mathcal{V}}
        = 
        \min_{\hat{p} \in \ProbDist(\hat{X})} \diamonddist{\sum_{x \in \hat{X}} \hat{p}(x) \mathcal{U}_x}{\mathcal{V}}. 
    \end{equation}
    Moreover, Algorithm~\ref{algo: sub prob synth} necessarily terminates in finite time. 
\end{prop}
\begin{proof}
    If Algorithm~\ref{algo: sub prob synth} terminates at Step~(c), then $\Ball{\delta}{V} = \UC(2)$ for any $t \in \N$, and the claim follows immediately from $\CliffordTOne_{\Tcount \leq t} \cap \Ball{\delta}{V} = \CliffordTOne_{\Tcount \leq t}$. 
    Otherwise, it holds trivially by Lemma~\ref{lem: min_X equal min_hatX}. 
    Note that, by the termination condition of Step~(c), we have $\delta \in (0, \frac12]$ at Step~(d), and therefore the algorithm of Lemma~\ref{lem: covering verification algorithm} can be applied. 
    The termination follows immediately from Proposition~\ref{prop: correctness divide and conquer sub dete synth}. 
\end{proof}

We analyze the computational cost and the outputs of Algorithm~\ref{algo: sub prob synth}. 
The following lemma provides an upper bound on the probability that Step~(d) is not passed.
\begin{lem}
\label{lem: bound of probability of failure of delta-covering}
    It holds that
    \begin{equation}
        \Pr \left[ \text{$\CliffordTOne_{\Tcount \leq t} \cap \mathbf{B}_{2\delta}(V)$ isn't a $\delta$-covering of $\mathbf{B}_\delta(V)$} \right] = O\left( \frac{t^2}{2^t \delta^3} \right) \quad \text{as $1/\delta,\; t \rightarrow \infty$}.
    \end{equation}
\end{lem}

\begin{proof}
    Since it is clear that
    \begin{equation}
        \Pr \left[ \text{$\CliffordTOne_{\Tcount \leq t} \cap \mathbf{B}_{2\delta}(V)$ isn't a $\delta$-covering of $\mathbf{B}_\delta(V)$} \right]
        = 
        \Pr \left[ \text{$\CliffordTOne_{\Tcount \leq t}$ isn't a $\delta$-covering of $\mathbf{B}_\delta(V)$} \right],
    \end{equation}
    we analyze the right-hand side instead of the left-hand side.
    Let $\{ U_x \}_{x \in X} = \CliffordTOne_{\leq t}$, let $\{ U_y \}_{y \in Y}$ be a finite set of single-qubit unitaries, and $c_1, c_2$ be positive real numbers. 
    Assume that 
    \begin{enumerate}
        \item $c_1 + c_2 \leq 1$ holds and,
        \item $\{ U_y \}_{y \in Y}$ is a $c_1\delta$-covering of $\mathbf{B}_\delta(I_2)$. 
    \end{enumerate}
    Under these assumptions, we first prove that 
    \begin{equation}
    \label{eq: prove of bound of probability of delta-covering 1}
        V \in \bigcap_{y \in Y} \left( \bigcup_{x \in X} \mathbf{B}_{c_2\delta}(U_y^\dag U_x) \right) 
        \ \Longrightarrow \ 
        \text{$\{ U_x \}_{x \in X}$ is a $\delta$-covering of $\mathbf{B}_\delta(V)$}.
    \end{equation}
    By the bi-invariance of the diamond distance, we can assume that $\{ U_y V \}_{y \in Y}$ is a $c_1 \delta$-covering of $\mathbf{B}_\delta(V)$. 
    Also, the condition on the left-hand side of Eq~\eqref{eq: prove of bound of probability of delta-covering 1} can be rewritten as
    \begin{align}
        V \in \bigcap_{y \in Y} \left( \bigcup_{x \in X} \mathbf{B}_{c_2\delta}(U_y^\dag U_x) \right) 
        &\ \Longleftrightarrow \
        \forall y \in Y, \ \exists x \in X, \ \diamonddist{\mathcal{V}}{\mathcal{U}_y^\dag \mathcal{U}_x} < c_2 \delta \\
        \label{eq: prove of bound of probability of delta-covering 2}
        &\ \Longleftrightarrow \ 
        \forall y \in Y, \ \exists x \in X, \ \diamonddist{\mathcal{U}_y \mathcal{V}}{\mathcal{U}_x}  < c_2 \delta. 
    \end{align}
    For any $W \in \mathbf{B}_\delta(V)$, the second assumption guarantees the existence of $y \in Y$ such that $\diamonddist{\mathcal{W}}{\mathcal{U}_y \mathcal{V}} < c_1 \delta$. Moreover, there exists $x \in X$ such that $\diamonddist{\mathcal{U}_y \mathcal{V}}{\mathcal{U}_x} < c_2\delta$ by Eq~\eqref{eq: prove of bound of probability of delta-covering 2}. 
    By the triangle inequality and the first assumption, the above-chosen $x$ and $y$ satisfy the following inequality:
    \begin{equation}
        \diamonddist{\mathcal{W}}{\mathcal{U}_x} \leq \diamonddist{\mathcal{W}}{\mathcal{U}_y\mathcal{V}} + \diamonddist{\mathcal{U}_y \mathcal{V}}{\mathcal{U}_x} < (c_1 + c_2)\delta \leq \delta.
    \end{equation}
    This fact implies Eq~\eqref{eq: prove of bound of probability of delta-covering 1}. 
    By combining this result with Lemma~\ref{lem: Clifford+T covering bound}, we have the following bound:
    \begin{align}
        \Pr \left[ \text{$\CliffordTOne_{\Tcount \leq t}$ isn't a $\delta$-covering of $\mathbf{B}_\delta(V)$} \right] 
        &\leq \Pr \left[ V \notin \bigcap_{y \in Y} \bigcup_{x \in X} \mathbf{B}_{c_2\delta}(U_y^\dag U_x) \right] \\
        &= \Pr \left[ \bigvee_{y \in Y} V \notin \bigcup_{x \in X} \mathbf{B}_{c_2\delta}(U_y^\dag U_x) \right] \\
        &\leq \sum_{y \in Y} \Pr \left[ V \notin \bigcup_{x \in X} \mathbf{B}_{c_2\delta}(U_y^\dag U_x) \right] \\
        &= \sum_{y \in Y} \Pr \left[ V \notin \bigcup_{x \in X} \mathbf{B}_{c_2\delta}(U_x) \right] \\
        &= O \left( |Y| \times \frac{t^2}{2^t (c_2\delta)^3} \right).
    \end{align}
    Then, the lemma holds if there exist positive real numbers $c_1, c_2$ and a finite set of single-qubit unitaries $\{ U_y \}_{y \in Y}$ that satisfy the above two assumptions and, in addition, $|Y| = O(1)$ and $c_2 = \Omega(1)$. 
    By setting $c_1 = c_2 = 0.5$,  we now show that such $\{ U_y \}_{y \in Y}$ exists.
    If the metric measure space $(\UC(2), d_\diamond, \mu_{\UC(2)})$ is doubling-metric \cite{heinonen2001lectures}, the claim follows immediately. 
    A straightforward measure-theoretic calculation shows that there exists a constant $M > 0$ such that
    \begin{equation}
        \frac{\muUCtwo \Bigl( \mathbf{B}_r(W) \Bigr)}{\mu_{\UC(2)} \left( \mathbf{B}_{\frac{r}{2}}(W) \right)} \leq M \quad \text{for any $W \in \UC(2)$ and any $r > 0$}. 
    \end{equation}
    Therefore, because $(\UC(2),\ d_\diamond,\ \muUCtwo)$ is doubling-measure, we can state that $(\UC(2),\ d_\diamond,\ \mu_{\UC(2)})$ is also doubling-metric \cite{heinonen2001lectures}. 
\end{proof}

\begin{prop}[Analysis of Algorithm~\ref{algo: sub prob synth}]
\label{prop: analysis sub prob synth}
    Let $\delta_t(V)$ denote the value of $\delta$ at the termination of Algorithm~\ref{algo: sub prob synth}, let $N_t(V) = \left| \CliffordTOne_{\Tcount \leq t} \cap \Ball{\delta_t(V)}{V} \right|$, and let $\CostAlg{\ref*{algo: sub prob synth}}{t}(V)$ be the number of arithmetic operations of Algorithm~\ref{algo: sub prob synth}. 
    Then, we have the following bounds:
    \begin{enumerate}[label=\textbf{\arabic*.}]
        \item for any $\gamma > 0$, $\delta_t = \Owhp(2^{-t/3 + \gamma t})$ as $2^t \rightarrow \infty$,
        \item for any $\gamma > 0$, $N_t = \Owhp(2^{\gamma t})$ as $2^t \rightarrow \infty$,
        \item for any $\gamma > 0$, $\CostAlg{\ref*{algo: sub prob synth}}{t} = \Owhp\left(2^{t/6 + \gamma t}\right)$ as $2^t \rightarrow \infty$. 
    \end{enumerate}
\end{prop}

\begin{proof}
    We first analyze the $\delta_t(V)$. 
    By Lemma~\ref{lem: bound of probability of failure of delta-covering}, we can see that
    \begin{equation}
    \label{eq: bound of probability of delta > alpha}
        \Pr \left[ \delta_t(V) > \alpha \right] \leq \Pr\Bigl[ \text{$\CliffordTOne_{\Tcount \leq t} \cap \Ball{2\alpha}{V}$ isn't a $\delta$-covering of $\Ball{\alpha}{V}$} \Bigr] = O\left( \frac{t^2}{2^t \alpha^3} \right)
    \end{equation}
    as $t,\alpha \rightarrow \infty$. 
    Consequently, for any $\gamma > 0$, it holds that 
    \begin{equation}
        \Pr[ \delta_t(V) > 2^{(-1 + \gamma)t/3} ] = O\left( 2^{-\gamma t}\poly(t) \right) \quad \text{as $2^t \rightarrow \infty$}. 
    \end{equation}
    We complete the proof of the first bound. 
    
    Next, we prove the second bound. 
    By the above discussion, we have
    \begin{align}
        \Pr[\left| \CliffordTOne_{\Tcount \leq t} \cap \Ball{\delta_t(V)}{V} \right| > \left| \CliffordTOne_{\Tcount \leq t} \cap \Ball{2^{(-1 + \gamma_1)t/3}}{V} \right|] &\leq \Pr[\delta_t(V) > 2^{(-1 + \gamma_1)t / 3}] \\
        &= O(2^{-\gamma_1 t}) \quad \text{as $2^t \rightarrow \infty$}
    \end{align}
    for any $\gamma_1 > 0$. 
    Therefore, it holds that
    \begin{equation}
        N_t = \Owhp\Bigl( \left| \CliffordTOne_{\Tcount \leq t} \cap \mathbf{B}_{2^{(-1 + \gamma_1)t/3}} \right| \Bigr) \quad \text{as $2^t \rightarrow \infty$}. 
    \end{equation}
    By Markov's inequality, for any $\gamma_2 > 0$, we obtain
    \begin{align}
        \Pr[ \left| \CliffordTOne_{\Tcount \leq t} \cap \Ball{2^{(-1 + \gamma_1)t / 3}}{V} \right| > 2^{(\gamma_1 + \gamma_2) t}]  
        &= 
        \Pr[\sum_{U \in \CliffordTOne_{\Tcount \leq t}} 1_{\Ball{2^{(-1 + \gamma_1)t / 3}}{U}}(V) > 2^{(\gamma_1 + \gamma_2) t}] \\
        &\leq
        2^{-(\gamma_1 + \gamma_2) t} \times \sum_{U \in \CliffordTOne_{\Tcount \leq t}} \Pr[V \in \Ball{2^{(-1 + \gamma_1)t / 3}}{U}] \\
        &= O\left(2^{-(\gamma_1 + \gamma_2) t} \times 2^t \times 2^{(-1 + \gamma_1)t} \right) \\
        &= O\left( 2^{-\gamma_2 t} \right) \quad \text{as $2^t \rightarrow \infty$}. 
    \end{align}
    Thus, we complete the proof of the second bound. 

    Finally, we prove the third bound. 
    Let $K_t(V)$ be the number of iterations of Steps~(b)-(d) in Algorithm~\ref{algo: sub prob synth}. 
    Then, it is easy to see that
    \begin{equation}
    \label{eq: cost = sum cost + |CT cap B|^4}
        \CostAlg{\ref*{algo: sub prob synth}}{t} = O\left( \sum_{i = 1}^{K_t} \left( \CostAlg{\ref*{algo: divide and conquer sub dete synth}}{t, 2\delta_t^{(i)}} + \left(N_t^{(i)}\right)^4 \right) \right) \quad \text{where }
        \begin{cases}
            \delta_t^{(i)} = 2^{-t/3 + c_1 + c_2 i} \\
            N_t^{(i)}(V) = \left| \CliffordTOne_{\Tcount \leq t} \cap \Ball{2\delta_t^{(i)}}{V} \right|
        \end{cases}
    \end{equation}
    By Eq.~\eqref{eq: bound of probability of delta > alpha}, for any $\gamma_1 > 0$, we have
    \begin{align}
        &\Pr[\sum_{i = 1}^{K_t(V)} \left( \CostAlg{\ref*{algo: divide and conquer sub dete synth}}{t, 2\delta_t^{(i)}}(V) + \left(N_t^{(i)}(V)\right)^4 \right) >  \sum_{i = 1}^{\gamma_1 t} \left( \CostAlg{\ref*{algo: divide and conquer sub dete synth}}{t, 2\delta_t^{(i)}}(V) + \left(N_t^{(i)}(V)\right)^4 \right)] \\
        = &\Pr[K_t(V) > \gamma_1 t] \\
        \leq &\Pr[\delta_t(V) > 2^{(-1 + c_1 + c_2 \gamma_1)t / 3}] \\
        = &O\left( 2^{(-c_1 - c_2\gamma_1)t} \poly(t) \right) \quad \text{as $2^t \rightarrow \infty$}.
    \end{align}
    Therefore, it holds that
    \begin{equation}
        \label{eq: cost + N = Owhp(cost + N)}
        \sum_{i = 1}^{K_t} \left( \CostAlg{\ref*{algo: divide and conquer sub dete synth}}{t, 2\delta_t^{(i)}} + \left(N_t^{(i)}\right)^4 \right)
        = 
        \Owhp\left(\sum_{i = 1}^{\gamma_1 t} \left( \CostAlg{\ref*{algo: divide and conquer sub dete synth}}{t, 2\delta_t^{(i)}} + \left(N_t^{(i)}\right)^4 \right) \right) \quad \text{as $2^t \rightarrow \infty$}. 
    \end{equation}
    We analyze the first term of R.H.S of Eq~\eqref{eq: cost + N = Owhp(cost + N)}. 
    By Markov's inequality and Proposition~\ref{prop: complexity divide and conquer sub dete synth}, for any $\gamma_2 > 0$, we have
    \begin{align}
        \Pr[\sum_{i=1}^{\gamma_1 t} \CostAlg{\ref*{algo: divide and conquer sub dete synth}}{t, 2\delta_t^{(i)}}(V) > 2^{(1 +5\gamma_1 +6\gamma_2)t/6}] &\leq 2^{-(1 +5\gamma_1 +6\gamma_2)t/6} \sum_{i=1}^{\gamma_1 t} \Expect\left[\CostAlg{\ref*{algo: divide and conquer sub dete synth}}{t, 2\delta_t^{(i)}} \right] \\
        &= O\left( 2^{-(1 +5\gamma_1 +6\gamma_2)t/6} \times 2^{(1 + 5\gamma_1)t/6} \polylog(t) \right) \\
        &= O\left( 2^{-\gamma_2 t}  \right) \quad \text{as $2^t \rightarrow \infty$}.
    \end{align}
    Therefore, it holds that
    \begin{equation}
        \sum_{i=1}^{\gamma_1 t} \CostAlg{\ref*{algo: divide and conquer sub dete synth}}{t, 2\delta_t^{(i)}} = \Owhp\left( 2^{(1 + 5\gamma_1 + 6\gamma_2)t/6} \right) \quad \text{as $2^t \rightarrow \infty$}. 
    \end{equation}
    Next, we analyze the second term of R.H.S of Eq~\eqref{eq: cost + N = Owhp(cost + N)}. 
    By Markov's inequality, for any $\gamma_3 > 0$, we have
    \begin{align}
        \Pr[\sum_{i=1}^{\gamma_1 t} \left(N_t^{(i)}(V) \right)^4 > 2^{4(\gamma_1 + \gamma_3)t} ] &\leq \Pr[\left(\sum_{i=1}^{\gamma_1 t} N_t^{(i)}(V) \right)^4 > 2^{4(\gamma_1 + \gamma_3)t}] \\
        &= \Pr[\sum_{i=1}^{\gamma_1 t} \sum_{U \in \CliffordTOne_{\Tcount \leq t}} 1_{\Ball{2\delta_t^{(i)}}{U}}(V)  > 2^{(\gamma_1 + \gamma_3)t}] \\
        &\leq 2^{-(\gamma_1 + \gamma_3)t} \times \sum_{i=1}^{\gamma_1 t} \sum_{U \in \CliffordTOne_{\Tcount \leq t}} \Pr[V \in \Ball{2\delta_t^{(i)}}{U}] \\
        &= O\left( 2^{-(\gamma_1 + \gamma_3)t} \times 2^{\gamma_1 t} \right) \\
        &= O\left( 2^{-\gamma_3 t} \right) \quad \text{as $2^t \rightarrow \infty$}. 
    \end{align}
    Therefore, it holds that
    \begin{equation}
        \sum_{i=1}^{\gamma_1 t} \left(N_t^{(i)} \right)^4 = \Owhp\left( 2^{4(\gamma_1 + \gamma_3)t}  \right) \quad \text{as $2^t \rightarrow \infty$}.
    \end{equation}
    These results imply the third bound.
\end{proof}

\subsection{Overall algorithm}

\begin{algo}[$T$-optimal Probabilistic Synthesis]
\label{algo: prob synth}
    Given a single-qubit unitary $V$ and a precision $\varepsilon > 0$.
    \begin{enumerate}
        \item If $\varepsilon > 1$, output $\{ I_2 \}$ and terminate the algorithm. 
        \item Set $t = 0$. 
        \item Run Algorithm~\ref{algo: sub prob synth} with input $(V, t)$ and denote its output by $\{U_x\}_{x \in \hat{X}}$. 
        \item Compute the optimal distance $\varepsilon^* = \min_{\hat{p} \in \ProbDist(\hat{X})} \diamonddist{\sum_{x \in \hat{X}}\hat{p}(x)\mathcal{U}_x}{\mathcal{V}}$ and its optimal probability distribution $\hat{p}$ throughout the SDP of Eq.~\eqref{eq: SDP compute optimal distribution}. 
        \item If $\varepsilon^* < \varepsilon$, output $\{U_x\}_{x \in \hat{X}}$ and $\hat{p}$. 
        Otherwise, update $t \leftarrow t+1$ and repeat Step~(c).
    \end{enumerate}
\end{algo}

\begin{prop}[Correctness and Termination of Algorithm~\ref{algo: prob synth}]
\label{prop: correctness of prob synth}
    Algorithm~\ref{algo: prob synth} necessarily outputs a solution to Problem~\ref{prob: prob synth} in finite time.
\end{prop}
\begin{proof}
    It follows immediately from Proposition~\ref{prop: correctness of sub prob synth}. 
\end{proof}

\begin{thm}[Complexity of Algorithm~\ref{algo: prob synth}]
\label{thm: complexity prob synth}
    Let $\TimeAlg{\ref*{algo: prob synth}}{\varepsilon}(V)$ be the runtime of Algorithm~\ref{algo: prob synth}. 
    For any $\gamma > 0$, it holds that
    \begin{equation}
        \TimeAlg{\ref*{algo: prob synth}}{\varepsilon} = \Owhp(\varepsilon^{-1/4 - \gamma}) \quad \text{as $1/\varepsilon \rightarrow \infty$}.
    \end{equation}
\end{thm}

\begin{proof}
    Let $\mixTcount_{\varepsilon}(V)$ denote the maximum value of $t$ in Algorithm~\ref{algo: prob synth}, and let $N_t(V)$ denote the cardinality of $\hat{X}$ at Step~(c). 
    It is easy to see that 
    \begin{equation}
        \TimeAlg{\ref*{algo: prob synth}}{\varepsilon} = O\left( \sum_{t=0}^{\mixTcount_\varepsilon} \left( \CostAlg{\ref*{algo: sub prob synth}}{t} + (N_t)^c \right) \times \polylog(1/\varepsilon) \right)  \quad \text{as $1/\varepsilon \rightarrow \infty$}
    \end{equation}
    some $c > 0$ independent of $V$. 
    Note that each arithmetic operation is performed with $O(\log(1/\varepsilon))$-bit precision. 
    From Lemma~\ref{lem: min_X equal min_hatX} and Lemma~\ref{lem: bound of probability of failure of delta-covering}, we have the following bound with respect to $\mixTcount_{\varepsilon}$:
    \begin{align}
        \Pr\left[ \mixTcount_\varepsilon(V) > \alpha \right] &\leq \Pr\Bigl[ \text{$\CliffordTOne_{\Tcount \leq \alpha} \cap B_{2\sqrt{\varepsilon}}(V)$ isn't a $\sqrt{\varepsilon}$-covering of $B_{\sqrt{\varepsilon}}(V)$} \Bigr] = O\left( \frac{\alpha^2}{2^\alpha \varepsilon^{3/2}} \right)
    \end{align} 
    as $1/\varepsilon,\; \alpha \rightarrow \infty$. 
    Thus, we have
    \begin{align}
        &\Pr[\sum_{t=0}^{\mixTcount_\varepsilon(V)} \left( \CostAlg{\ref*{algo: sub prob synth}}{t}(V) + N_t(V)^c \right) > \sum_{t=0}^{(1.5+\gamma_1)\log_2(1/\varepsilon)} \left( \CostAlg{\ref*{algo: sub prob synth}}{t}(V) + N_t(V)^c \right) ] \\ 
        = &\Pr[\mixTcount_\varepsilon(V) > (1.5 + \gamma_1)\log_2(1/\varepsilon)] \\
        = &O\left( \varepsilon^{\gamma_1} \polylog(1/\varepsilon) \right) \quad \text{as $1/\varepsilon \rightarrow \infty$}. 
    \end{align}
    Therefore, it holds that
    \begin{equation}
        \sum_{t=0}^{\mixTcount_\varepsilon} \left( \CostAlg{\ref*{algo: sub prob synth}}{t} + N_t^c \right) = \Owhp\left( \sum_{t=0}^{(1.5 + \gamma_1)\log_2(1/\varepsilon)} \left( \CostAlg{\ref*{algo: sub prob synth}}{t} + N_t^c \right) \right) \quad \text{as $1/\varepsilon \rightarrow \infty$}.
    \end{equation}
    By Proposition~\ref{prop: analysis sub prob synth}, for any $\gamma_2 > 0$, there exist $\alpha, C_\alpha, c_\alpha, t_\alpha > 0$ such that
    \begin{align}
        \Pr[\sum_{t = t_\alpha}^{(1.5 + \gamma_1)\log_2(1/\varepsilon)} \CostAlg{\ref*{algo: sub prob synth}}{t}(V) > \alpha \sum_{t = t_\alpha}^{(1.5 + \gamma_1)\log_2(1/\varepsilon)}  2^{t/6 + \gamma_2 t}] 
        &\leq 
        \Pr[\bigvee_{t=t_\alpha}^{(1.5 + \gamma_1)\log_2(1/\varepsilon)} \CostAlg{\ref*{algo: sub prob synth}}{t}(V) > \alpha 2^{t/6 + \gamma_2 t}] \\
        &\leq 
        \sum_{t = t_\alpha}^{(1.5 + \gamma_1)\log_2(1/\varepsilon)} \Pr[\CostAlg{\ref*{algo: sub prob synth}}{t} > \alpha 2^{t/6 + \gamma_2 t}] \\
        &\leq \sum_{t=t_\alpha}^{(1.5 + \gamma_1)\log_2(1/\varepsilon)} C_\alpha 2^{-c_{\alpha} t} \\
        &= O(\poly(\varepsilon)) \quad \text{as $1/\varepsilon \rightarrow \infty$}. 
    \end{align}
    Furthermore, since Algorithm~\ref{algo: sub prob synth} halts in finite time, there exists a constant $M > 0$ such that
    \begin{equation}
        \sum_{t=0}^{t_\alpha-1} \CostAlg{\ref*{algo: sub prob synth}}{t}(V) < M
    \end{equation}
    for all $V \in \UC(2)$. 
    Hence, it follows that
    \begin{alignat}{2}
        \sum_{t=0}^{(1.5 + \gamma_1)\log_2(1/\varepsilon)} \CostAlg{\ref*{algo: sub prob synth}}{t} &= 
        \Owhp\left( \sum_{t=0}^{(1.5 + \gamma_1)\log_2(1/\varepsilon)} 2^{t/6 + \gamma_2t} \right) &\quad \text{as $1/\varepsilon \rightarrow \infty$} \\
        &= 
        \Owhp(\varepsilon^{-1/4 - \gamma_1/6 - 1.5\gamma_2 - \gamma_1 \gamma_2}) &\quad \text{as $1/\varepsilon \rightarrow \infty$}. 
    \end{alignat}
    In the same manner, for any $\gamma_3 > 0$, we obtain
    \begin{alignat}{2}
        \sum_{t=0}^{(1.5 + \gamma_1)\log_2(1/\varepsilon)} N_t^c &= \Owhp\left( \sum_{t=0}^{(1.5 + \gamma_1)\log_2(1/\varepsilon)} 2^{c\gamma_3 t} \right) &\quad \text{as $1/\varepsilon \rightarrow \infty$} \\
        &= \Owhp\left( \varepsilon^{-c\gamma_3(1.5 + \gamma_1)} \right) &\quad \text{as $1/\varepsilon \rightarrow \infty$}. 
    \end{alignat}
    Thus, we complete the proof. 
\end{proof}

\begin{thm}[Bound on the $T$-count for Probabilistic Synthesis]
\label{thm: upper bound on T-count for Probabilistic}
    Let $\mixTcount_\varepsilon(V)$ be the $T$-count of a solution to Problem~\ref{prob: prob synth}. 
    It holds that 
    \begin{equation}
        \mixTcount_\varepsilon = 1.5\log_2(1/\varepsilon) + \owhp(\log_2(1/\varepsilon)) \quad \text{as $1/\varepsilon \rightarrow \infty$}.
    \end{equation}
\end{thm}

\begin{proof}
    The proof of this theorem is contained in the proof of Theorem~\ref{thm: complexity prob synth}. 
    Next, we prove the lower bound. 
    Lemma~4.1 of \cite{akibue2024probabilistic} shows that, for a finite set of single-qubit unitaries $\{U_x\}_{x \in X}$, it holds that
    \begin{equation}
        \min_{x \in X} \diamonddist{\mathcal{U}_x}{\mathcal{V}}^2 \leq \min_{p \in \ProbDist(X)}\diamonddist{\sum_{x \in X}p(x)\mathcal{U}_x}{\mathcal{V}}. 
    \end{equation}
    In order for $\min_{p \in \ProbDist}\diamonddist{\sum_{x \in X}p(x)\mathcal{U}_x}{\mathcal{V}} < \varepsilon$ to hold, it is necessary that there exists some $x \in X$ such that $\diamonddist{\mathcal{U}_x}{\mathcal{V}} < \sqrt{\varepsilon}$. 
    Therefore, we have
    \begin{align}
        \Pr[\mixTcount_\varepsilon(V) \leq (1.5 - \gamma)\log_2(1/\varepsilon)] &\leq \Pr[V \in \bigcup_{U \in \CliffordTOne_{\Tcount \leq (1.5-\gamma)\log_2(1/\varepsilon)}} \Ball{\sqrt{\varepsilon}}{U}] \\
        &= \sum_{U \in \CliffordTOne_{\Tcount \leq (1.5-\gamma)\log_2(1/\varepsilon)}} \Pr[V \in \Ball{\sqrt{\varepsilon}}{U}] \\
        &= O(\varepsilon^{-1.5 + \gamma}) \times O(\varepsilon^{1.5}) = O(\varepsilon^\gamma). 
    \end{align}
\end{proof}

\section{Numerical experiment \label{sec: Numerical Experiment}}
\begin{figure}[h]
    \centering
    \includegraphics[width=0.8 \linewidth]{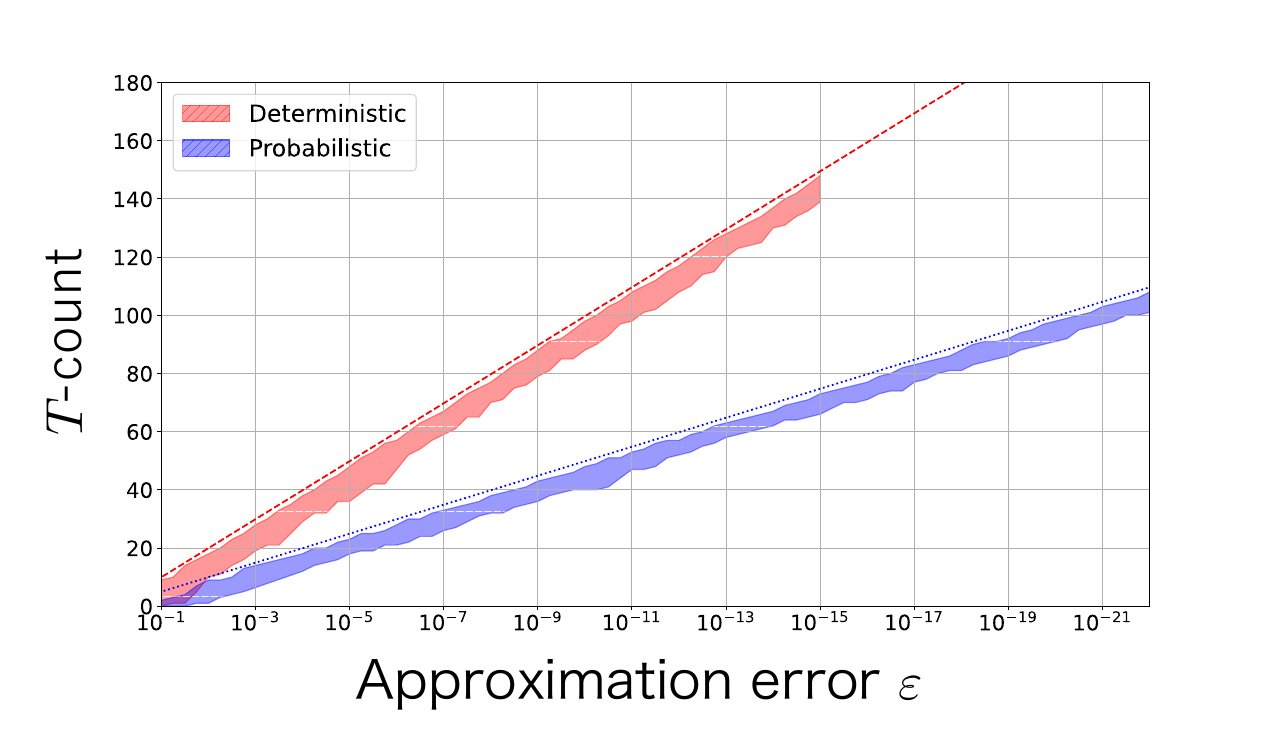}
    \caption{The $T$-count required for Algorithms~\ref{algo: dete synth} and \ref{algo: prob synth}. Each colored region represents the range between the minimum and maximum $T$-counts. Also, the purple dashed line represents $y = 3\log_2(1/\varepsilon)$, and the red dashed line represents $y = 1.5\log_2(1/\varepsilon)$.}
    \label{fig: t-count}
\end{figure}

\begin{figure}[h]
  \begin{minipage}[b]{0.5\linewidth}
    \centering
    \includegraphics[keepaspectratio, scale=0.5]{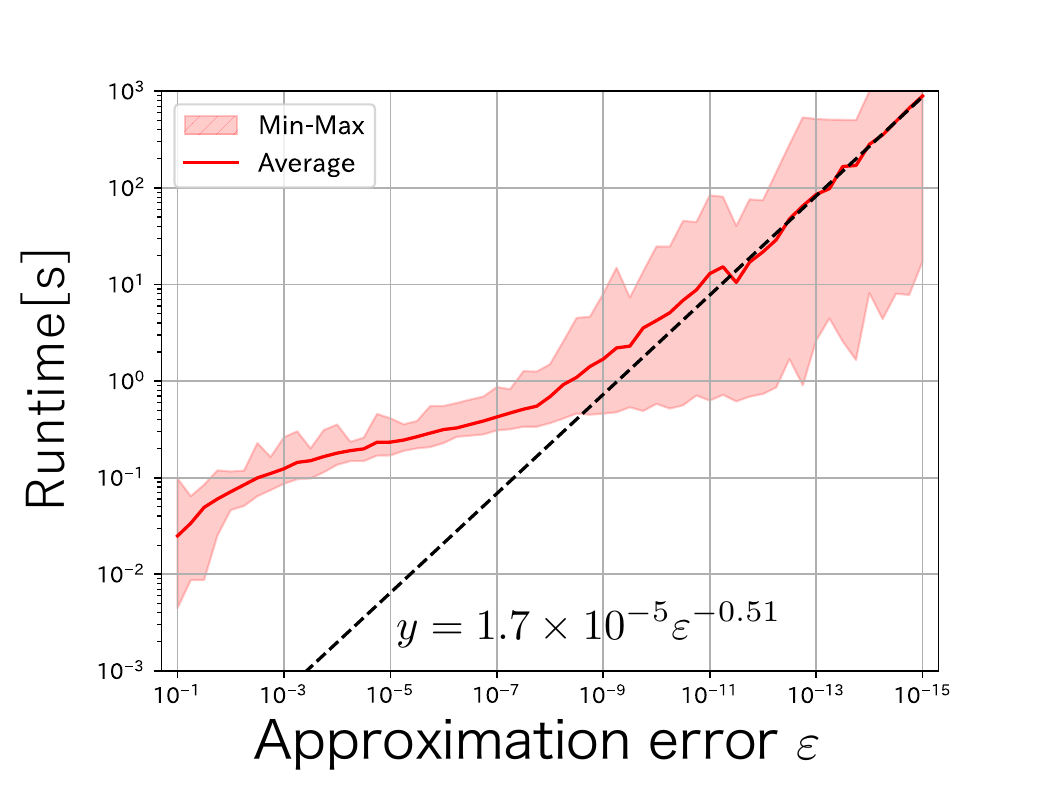}
    \subcaption{Deterministic}
  \end{minipage}
  \begin{minipage}[b]{0.5\linewidth}
    \centering
    \includegraphics[keepaspectratio, scale=0.5]{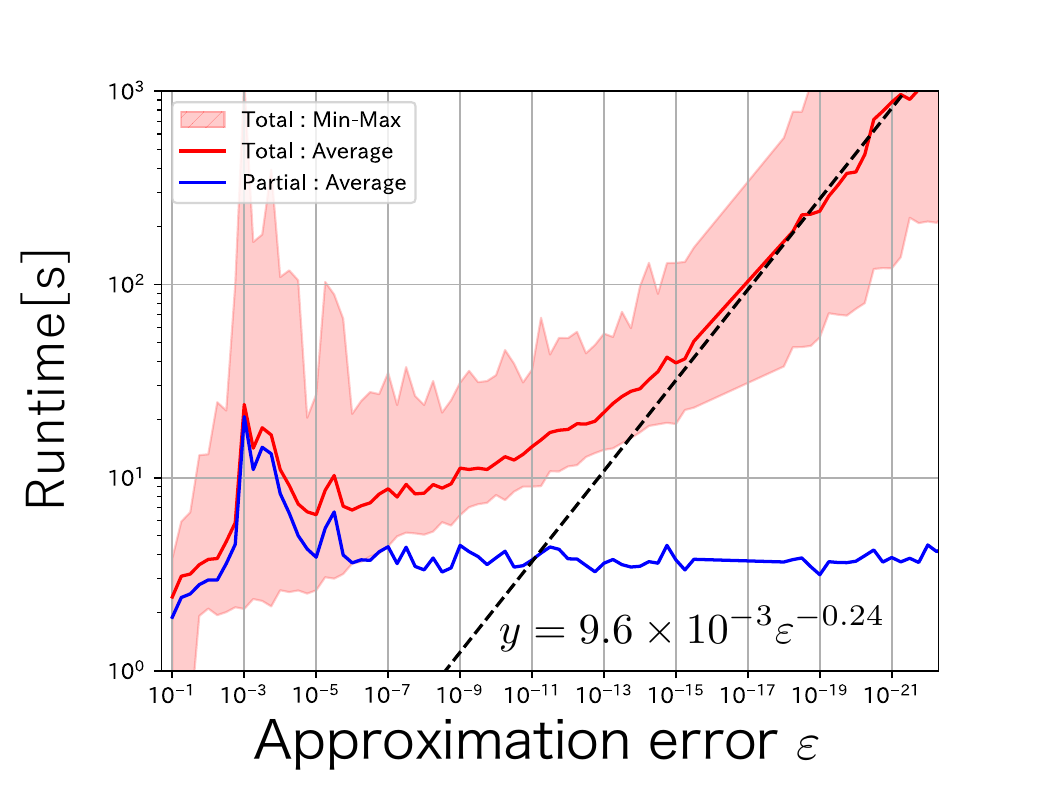}
    \subcaption{Probabilistic}
  \end{minipage}
  \caption{Runtime of Algorithms~\ref{algo: dete synth} and \ref{algo: prob synth}. The black dashed lines represent polynomial regressions fitted using the rightmost ten data points around each average. The blue line in (b) denotes the average runtime of the SDP in step~(d) and the algorithm in Lemma~\ref{lem: covering verification algorithm}.}
  \label{fig: runtime}
\end{figure}

We evaluate the performance of the Clifford+$T$ synthesis algorithms introduced in Sections~\ref{sec: Deterministic Synthesis} and \ref{sec: Probabilistic Synthesis}.
Both algorithms are implemented in C++ and benchmarked on Intel Xeon Platinum 9242. 
In our numerical experiments,  we sample 100 single-qubit unitaries $V \in \UC(2)$ from the Haar measure $\mu_{\UC(2)}$. 
For each target unitary $V$ and several approximation errors $\varepsilon$, we run Algorithm~\ref{algo: dete synth} and Algorithm~\ref{algo: prob synth}, and evaluate the performance of the $T$-count and runtime for each synthesis algorithm. 
The results are shown in Fig.~\ref{fig: t-count} and Fig~\ref{fig: runtime}. 
These results provide a numerical validation of the theoretical analyses conducted in Sections~\ref{sec: Deterministic Synthesis} and \ref{sec: Probabilistic Synthesis}, and show that the two proposed synthesis algorithm can be executed to $\varepsilon \approx 10^{-15}$ and $\varepsilon \approx 10^{-22}$, respectively, within roughly 15 minutes. 
The achieved approximation errors are sufficient for many practical quantum computing tasks, and we believe that our proposed synthesis algorithms are highly useful in such settings. 

\section{Conclusion}
We proposed two ancilla-free Clifford+$T$ synthesis algorithms for single-qubit unitaries that are optimal with respect to $T$-count. 
We showed that, with high probability, the deterministic synthesis algorithm proposed by Section~\ref{sec: Deterministic Synthesis} runs in time $\varepsilon^{-1/2 - \owhp(1)}$ and can approximate single-qubit unitaries with at most $3\log_2(1/\varepsilon) + \owhp(\log_2(1/\varepsilon))$. 
Also, we showed that, with high probability, the probabilistic synthesis algorithm proposed by Section~\ref{sec: Probabilistic Synthesis} runs in time $\varepsilon^{-1/4 - \owhp(1)}$ and can approximate single-qubit unitaries with at most $1.5\log_2(1/\varepsilon) + \owhp(\log_2(1/\varepsilon))$. 
These theoretical analyses don't rely on any conjecture. 
Moreover, we numerically confirmed the practicality of the proposed algorithms. 
The deterministic and probabilistic synthesis can run within feasible up to $\varepsilon \approx \varepsilon^{-15}$ and $\varepsilon \approx 10^{-22}$, respectively.

We have several interesting questions regarding this study. 
The first question concerns extending our algorithms and theoretical framework beyond the Clifford+$T$ gate set. 
Much of our analysis in this work relied on the number-theoretic properties of Clifford+$T$ gates, which played an essential role. 
Accordingly, it seems natural to explore extensions to other gate sets exhibiting similar number-theoretic structures, such as the Clifford+$V$ gate set considered in \cite{kliuchnikov2023shorter} and \cite{parzanchevski2018super}. 
We are also interested in how the runtime and gate complexity of such synthesis algorithms may be affected. 
The next question is whether Problems~\ref{prob: dete synth} and \ref{prob: prob synth} can be solved by algorithms running in polynomial time with respect to the input size, i.e., $\polylog(1/\varepsilon)$ time. 
Parzanchevski and Sarnak~\cite{parzanchevski2018super} show that an integer enumeration problem closely related to our Problem~\ref{prob: sub dete synth} is NP-complete. 
However, this result does not fully answer our question. 
If constructing such polynomial-time algorithms is indeed difficult, we are further interested in how much the optimality condition on the $T$-count must be relaxed in order to admit polynomial-time algorithms. 
For example, we ask whether there exists a polynomial-time algorithm that approximates a single-qubit unitary $V$ with $\Tcount_\varepsilon(V) + o(\log(1/\varepsilon))$ $T$-gates, where $\Tcount_\varepsilon(V)$ is the quantity defined in the proof of Theorem~\ref{thm: complexity dete synth}.

\section{Acknowledgements}
We thank Keisuke Fujii for useful discussions. 
H.M. is partially supported by MEXT Q-LEAP Grant No. JPMXS0120319794 and JST SPRING Grant No. JPMJSP2138. 
S.A. is partially supported by JST PRESTO Grant no.JPMJPR2111, JST Moonshot R\&D MILLENNIA Program (Grant no.JPMJMS2061), JPMXS0120319794, and CREST (Japan Science and Technology Agency) Grant no.JPMJCR2113.


\bibliographystyle{unsrt}
\bibliography{cite}

\newpage
\appendix

\section{Analysis of $r(m)$ \label{appendix: analysis of r(m)}}
In this section, let $K = \Q(\sqrt{2})$ and $\calO_K = \Z[\sqrt{2}]$.
This section aims to prove the following theorem.
\begin{thm}\label{thm: AppendixA: order of r(m)}
    For $m \in \calO_K$, let
    \[
        r_K(m) \coloneqq \#\{ (a,b,c,d)\in \Z[\sqrt{2}]^{4}\ |\ a^2 + b^2 + c^2 + d^2 = m\}.
    \]
    Then, we have
    \[
        r_K(m) \in O\left( N_{K/\Q}(m) (\log \log N_{K/\Q}(m))^2\right).
    \]
\end{thm}

The proof relies on the following variant of the four-square theorems for our $K$.
\begin{thm}[{\cite[Theorem 1.2]{doi:10.1142/S1793042124500039}}]\label{thm: AppendixA: explicit r(m)}
    If $m \in \Z[\sqrt{2}]\cap [0,\infty)$ is represented by the sum of four squares of elements in $\Z[\sqrt{2}]$, the number
    \[
        r_K(m) \coloneqq \#\{ (a,b,c,d)\in \Z[\sqrt{2}]^{4}\ |\ a^2 + b^2 + c^2 + d^2 = m\}
    \]
    is explicitly written as
    \begin{equation}\label{eq: explicit r(m)}
        r_K(m) = 8\sum_{(d)|(m)} N_{K/\Q}(d) - 6 \sum_{(2)|(d)|(m)}N_{K/\Q}(d) + 4 \sum_{(4)|(d)|(m)}N_{K/\Q}(d),
    \end{equation}
    where $N_{K/\Q}$ is the norm map.
\end{thm}
In \cite{doi:10.1142/S1793042124500039}, the same assertion for locally representable $m$ is proved.
Since we only need the order of $r_K(m)$, we focus on the case that $m$ is represented as a sum of four squares.

The following proposition is fundamental. We omit the proof of it.
\begin{prop}
    Let $p$ be a prime integer.
    Then, the following statements hold.
    \begin{enumerate}
        \item Suppose $p\equiv 3,5\mod 8$. Then, $\frakp \coloneqq p\calO_K$ is a prime ideal. Moreover, we have
        \[
            N_{K/\Q}(\frakp) = p^2.
        \]
        \item Suppose $p\equiv 1,7 \mod 8$. Then, the ideal $p\calO_K$ splits to a product of two distinct prime ideals $\frakp \overline{\frakp}$. Moreover, we have
        \[
            N_{K/\Q}(\frakp) = N_{K/\Q}(\overline{\frakp}) = p.
        \]
        \item Suppose $p=2$. Then, the ideal $2\calO_K$ is a square of a prime ideal $\frak{p}=\sqrt{2}\calO_K$.
        Moreover, we have
        \[
            N_{K/\Q}(\frakp) = p.
        \]
    \end{enumerate}
\end{prop}

\begin{proof}[Proof of Theorem~\ref{thm: AppendixA: order of r(m)}]
    For $m \in \calO_K$ representable as a sum of four squares of elements in $\calO_K$, since we have Theorem~\ref{thm: AppendixA: explicit r(m)}, it is enough to show that
    \[
        \sigma_{K/\Q}(m) \coloneqq \sum_{(d)|(m)} N_{K/\Q}(d)\in O\left( N_{K/\Q}(m) (\log \log N_{K/\Q}(m))^2\right)
    \]
    holds.
    Here, we note that the second and the third terms of Theorem~\ref{eq: explicit r(m)} are written as
    \begin{align}
        \sum_{(2)|(d)|(m)} N_{K/\Q}(d) &=
        \begin{cases}
            4 \sigma_{K/\Q}(m/2) &\text{if }2|m, \\
            0 & \text{otherwise},
        \end{cases}\\
        \sum_{(4)|(d)|(m)} N_{K/\Q}(d) &=
        \begin{cases}
            16 \sigma_{K/\Q}(m/4) &\text{if }4|m, \\
            0 & \text{otherwise}.
        \end{cases}
    \end{align}
    If $m$ is not representable as a sum of four squares of elements in $\calO_K$, we obviously have $r_{K/\Q}(m) = 0$.
    Thus, we may focus on the case that $m$ is represented as a sum of four squares of elements in $\calO_K$.
    
    For $m = a + b\sqrt{2} \in \calO_K$, we calculate $\sigma_{K/\Q}(m)$ using the prime number decomposition of $N_{K/\Q}(m) = a^2 - 2b^2$.
    Let $N_{K/\Q}(m) = \pm \prod_{i=1}^r p_i^{e_i'}$ be the prime number decomposition.
    Put
    \begin{align}
        A &\coloneqq \{ p_i \ |\ p_i \equiv 3,5 \mod 8\} = \{p_1,\ldots, p_s\}\\
        B &\coloneqq \{ p_i\ |\ p_i \equiv 1,7 \mod 8\} = \{ p_{s+1},\ldots, p_r\}.
    \end{align}
    For $p_i \in A$, let
    \[
        e_i \coloneqq \frac{1}{2} \ord_{p_i}(N_{K/\Q}(m))\ \in \Z.
    \]
    For $p_i \in B$, let $p_i\calO_K = \frakp_i \overline{\frakp_i}$ be the prime ideal decomposition, and let
    \begin{align}
        \overline{e_i} &\coloneqq \max\{e \in \Z\ |\ p_i^{e} \text{ divides }m \},\\
        e_i'' &\coloneqq \ord_{p_i}\left( \frac{N_{K/\Q}(m)}{p_i^{2\overline{e_i}}}\right),\\
        e_i &\coloneqq \overline{e_i} + e_i''.
    \end{align}
    Let
    \[
        e = \ord_{2}N_{K/\Q}(m).
    \]
    Then, if it is necessary, replacing $\frakp_i$ and $\overline{\frakp_i}$, the ideal $(m)$ is decomposed as
    \[
        (m) = \prod_{p_i \in A} (p_i)^{e_i} \cdot \prod_{p_i \in B} \frakp_i^{e_i} \overline{\frakp_i}^{\overline{e_i}} \cdot (\sqrt{2})^e.
    \] 
    The ideal $(d)$ divides $(m)$ if and only if $(d)$ is written as
    \[
        (d) = \prod_{p_i\in A}(p_i)^{g_i} \cdot \prod_{p_i \in B} \frakp_i^{g_i}\overline{\frakp_i}^{\overline{g_i}} \cdot (\sqrt{2})^{g}
    \]
    with
    \begin{align}
        0 &\leq g_i \leq e_i \text{ for }1\leq i \leq r,\\
        0 &\leq \overline{g_i} \leq \overline{e_i} \text{ for }s+1\leq i \leq r, \text{ and}\\
        0 &\leq g \leq e.
    \end{align}
    Hence, we obtain the equalities
    \begin{align}
        \sigma_{K/\Q}(m) =& \sum_{(d)|(m)} N_{K/\Q}(d)\\
        =& \prod_{1\leq i \leq s}\left( 1 + N_{K/\Q}(p_i)+ \cdots +  N_{K/\Q}(p_i)^{e_i}\right) \cdot
        \prod_{s+1\leq i \leq r}\left( 1 + N_{K/\Q}(\frakp_i)+ \cdots + N_{K/\Q}(\frakp_i)^{e_i}\right) \cdot\\
        & \prod_{s+1\leq i \leq r}\left( 1 + N_{K/\Q}(\overline{\frakp_i})+ \cdots + N_{K/\Q}(\overline{\frakp_i})^{\overline{e_i}}\right) \cdot
        \left( 1 + 2 + \cdots 2^e\right)\\
        =& \prod_{1\leq i \leq s} \frac{p_i^{2(e_i+1)-1}}{p_i^2 - 1} \cdot \prod_{1\leq i \leq s} \frac{(p_i^{e_i+1}-1)(p_i^{\overline{e_i}+1}-1)}{(p_i - 1)^2} \cdot (2^{e+1}-1).
    \end{align}
    
    On the other hand, we have
    \[
        \sigma(N_{K/\Q}(m)) = \prod_{p_i \in A} \frac{p_i^{2e_i+1}-1}{p_i -1 } \cdot \prod_{p_i \in B}\frac{p_i^{e_i + \overline{e_i}+ 1} - 1}{p_i - 1} \cdot (2^{e+1} - 1).
    \]
    Hence, we have the following inequalities
    \begin{align}
        \frac{\sigma_{K/\Q}(m)}{\sigma(N_{K/\Q}(m))}
        &= \prod_{p_i\in A} \frac{p_i^{e_i+\overline{e_i}+2} - p_i^{e_i + 1} - p_i^{\overline{e_i}+1} + 1}{(p_i^{e_i + \overline{e_i}+1}-1)(p_i - 1)} \cdot
        \prod_{p_i\in B}\frac{p_i^{2(e_i+1) - 1}}{(p_i^{2e_i + 1}-1)(p_i + 1)}\\
        &= \prod_{p_i\in A} \frac{p_i^{e_i+\overline{e_i}+2} - p_i^{e_i + 1} - p_i^{\overline{e_i}+1} + 1}{p_i^{e_i+\overline{e_i}+2} - p_i^{e_i + \overline{e_i} + 1} - p_i + 1} \cdot
        \prod_{p_i\in B}\frac{p_i^{2(e_i+1) - 1}}{p_i^{2e_i + 2} + p_i ^{2e_i + 1} - p_i + 1}\\
        &\leq \prod_{p_i \in A}\frac{p_i}{p_i - 1} \cdot \prod_{p_i \in B}\frac{p_i^{2e_i +2}}{p_i^{2e_i +2}}\\
        &\leq \prod_{p| N_{K/\Q}(m)} \frac{p}{p-1} = \frac{N_{K/\Q}(m)}{\varphi(N_{K/\Q}(m))} \label{eq: sigma leq sigma(N) N / phi}.
    \end{align}
    Let $\gamma$ be the Euler constant.
    The following well-known equalities due to Landau \cite[Satz in p.217]{Lan1909} and Gronwall \cite[Equation (22)]{Gron1913}
    \begin{align}
        & \liminf_{n\to \infty} \frac{\varphi(n) \log\log n}{n} = e^{-\gamma} \text{ and}\\
        & \limsup_{n\to\infty} \frac{\sigma(n)}{n\log\log n} = e^{\gamma}
    \end{align}
    implies that
    \begin{equation}\label{eq: order of sigma / phi}
        \frac{\sigma(n)}{\varphi(n)} \in O((\log\log n)^2).
    \end{equation}
    Our assertion follows from Eq.~\eqref{eq: sigma leq sigma(N) N / phi} and Eq.~\eqref{eq: order of sigma / phi}.
\end{proof}

\section{Proof of Lemma~\ref{lem: covering verification algorithm} \label{appendix: proof of covering verification algorithm}}

In this appendix, we will prove Lemma~\ref{lem: covering verification algorithm}. 
We first state the lemma that will be crucial for this proof. 
\begin{lem}
\label{lem: equivalence covering condition}
    Let $X$ be a topological space. 
    Let $S$ be a connected set $S \subseteq X$ and $\{A_i \}_{i \in I}$ be a finite family of open subsets of $X$. 
    Assume that $S \cap \left( \bigcup_{i \in I} A_i \right) \neq \emptyset$. 
    Then, it holds that the following two statements are equivalent:
    \begin{enumerate}
        \item[(1)] $S \subseteq \bigcup_{i \in I} A_i$,
        \item[(2)] $\forall j \in I$, $S \cap \partial A_j \subseteq \bigcup_{i \in I} A_j$.
    \end{enumerate}
\end{lem}

\begin{proof}
    We prove $(1) \Rightarrow (2)$.
    If $S \cap \partial A_j = \emptyset$, it holds trivially (2).
    When $S \cap \partial A_j \neq \emptyset$, let $x \in S \cap \partial A_j$, then $x \in S$.
    From (1), we see that $x \in \bigcup_{i \in I} A_i$.

    We next prove $(1) \Leftarrow (2)$. 
    First, we consider the case of $|I|=1$, i.e., $\{A_i\}_{i \in I} = \{A\}$.
    If $S \cap \partial A \neq \emptyset$, it doesn't hold (1) by the assumption that $A$ is an open set. 
    Therefore, we consider only $S \cap \partial A = \emptyset$.
    Then, we see
    \begin{equation}
        S \cap \mathrm{cl}(A) = S \cap (\partial A \cup A^\circ) = S \cap (\partial A \cup A) = (S \cap \partial A) \cup (S \cap A) = S\cap A.
    \end{equation}
    Since $\mathrm{cl}(A)$ is a closed set in $X$ and $A$ is an open set in $X$, $S \cap A$ is both an open set and a closed set in the relative topology on $S$.
    From the connectedness of $S$, any subset that is both open and closed in $S$ must be either $S$ itself or $\emptyset$.
    By the assumption of $S \cap A \neq \emptyset$, we have $S \cap A = S$, and prove (1).

    Next, let us consider the case $|I| > 1$.
    From the above discussion, if we define $A = \bigcup_{i \in I}A_i$, then the following holds:
    \begin{equation}
        S \cap \partial A \subseteq \bigcup_{i \in I} A_i \ \Longrightarrow \ S \subseteq \bigcup_{i \in I} A_i.
    \end{equation}
    Since $I$ is a finite set, we have
    \begin{equation}
        \partial A = \partial \left( \bigcup_{i \in I} A_i \right) \subseteq \bigcup_{i \in I} \partial A_i.
    \end{equation}
    Therefore, it holds that
    \begin{equation}
        (2) \ \Longleftrightarrow \ S \cap \left(\bigcup_{i \in I} \partial A_i \right) \subseteq \bigcup_{i \in I} A_i \ \Longrightarrow \ S \cap \partial \left( \bigcup_{i \in I} A_i \right) \subseteq \bigcup_{i \in I} A_i \ \Longrightarrow \ (1).
    \end{equation}
\end{proof}

The sketch of our proof is to recursively apply Lemma~\ref{lem: equivalence covering condition}, thereby reducing the original covering verification problem for an infinite set to that for a finite set, i.e., a finite number of membership verification problems. 
Note that, if $S \cap \left( \bigcup_{i \in I} A_i \right) \neq \emptyset$, then it is clear that $S \nsubseteq \bigcup_{i \in I} A_i$. 
By the isomorphism between $\UC(2)$ and $\widetilde{S}^3 \coloneqq S^3 / \{e,-e\}$, we may identify the covering verification problem on $\widetilde{S}^3$ with the following:
\begin{center}
    Given $v \in \widetilde{S}^3$, a finite set $\{u_i\}_{i \in I} \subseteq \widetilde{S}^3$ and $\delta \in (0, \frac12]$, determine whether $\Ball{\delta}{v} \subseteq \bigcup_{i \in I} \Ball{\delta}{u_i}$,
\end{center}
where the ball $\Ball{r}{x}$ centered at $u \in \widetilde{S}^3$ with radius $r > 0$ is defined as
\begin{equation}
    \left\{ u \in \widetilde{S}^3 \in \ | \   |u \cdot x| > \sqrt{1 - \min(1, \ r^2)}  \right\}. 
\end{equation}
We may, without loss of generality, assume that $u_i \neq u_j$ whenever $i \neq j$. 
This means that, for any choice of representatives, we have $u_i \not \perp u_j$.

We consider applying Lemma~\ref{lem: equivalence covering condition} to the original covering verification problem to reduce it to another covering verification problem.
It is easy to see that
\begin{align}
    \Ball{\delta}{v} 
    &= 
    \pi \left( 
        \{ x \in S^3 \ | \ u \cdot x > \sqrt{1 - \delta^2} \}
    \right)
\end{align}
where $\pi : S^3 \rightarrow \widetilde{S}^3$ is the quotient mapping. 
By the continuity of $\pi$, we can see that the set $\Ball{\delta}{v}$ is connected. 
Therefore, if it holds that
\begin{equation}
    \Ball{\delta}{v} \cap \left( \bigcup_{i \in I} \Ball{\delta}{u_i} \right) = \bigcup_{i \in I} \mathbf{B}_\delta(v) \cap \Ball{\delta}{u_i} \neq \emptyset,
\end{equation}
we can apply Lemma~\ref{lem: equivalence covering condition} and see that 
\begin{align}
    \mathbf{B}_\delta(v) \subseteq \bigcup_{i \in I} \Ball{\delta}{u_i})
    \Longleftrightarrow 
    \forall j \in I, \ \mathbf{B}_\delta(v) \cap \partial \mathbf{B}_\delta(u_j) \subseteq \bigcup_{i \in I_j} \Ball{\delta}{u_i},
\end{align}
where $I_j = I_j$. 
Note that, since $\Ball{\delta}{u_j}$ is a open set, it holds that $\Ball{\delta}{u_j} \cap \partial\Ball{\delta}{u_j} = \emptyset$. 
Moreover, it is possible to check whether $\Ball{\delta}{v} \cap \Ball{\delta}{u_j} $ is empty by performing some basic linear-algebraic calculations. 
Next, we consider the covering verification problem of
\begin{equation}
\label{eq: B dB <= cup B}
    \mathbf{B}_\delta(v) \cap \partial \mathbf{B}_\delta(u_j) \subseteq \bigcup_{i \in I_j} \Ball{\delta}{u_i}. 
\end{equation}
We can see that
\begin{align}
\label{eq: B cap dB}
    \mathbf{B}_{\delta}(v) \cap \partial \mathbf{B}_{\delta}(u_1) 
    &= 
    \pi \left( \left\{ x \in S^3 \ \middle| \ 
    \begin{array}{rrr}
        v \cdot x & > & \sqrt{1 - \delta^2} \\
        u_1 \cdot x & = & \sqrt{1 - \delta^2}
    \end{array} 
    \right\} 
    \cup 
    \left\{ x \in S^3 \ \middle| \ 
    \begin{array}{rrr}
        v \cdot x & > & \sqrt{1 - \delta^2} \\
        u_1 \cdot x & = & -\sqrt{1 - \delta^2}
    \end{array} 
    \right\} 
    \right) \\
    &= 
    \label{eq: B dB 3}
    \underbrace{
    \pi\left(
        \left\{ x \in S^3 \ \middle| \ 
        \begin{array}{rrr}
            v \cdot x & > & \sqrt{1 - \delta^2} \\
            u_1 \cdot x & = & \sqrt{1 - \delta^2}
        \end{array} 
        \right\} 
    \right)}_{\displaystyle \coloneqq S_j^+}
    \cup 
    \underbrace{
    \pi\left(
        \left\{ x \in S^3 \ \middle| \ 
        \begin{array}{rrr}
            v \cdot x & > & \sqrt{1 - \delta^2} \\
            u_1 \cdot x & = & -\sqrt{1 - \delta^2}
        \end{array} 
        \right\} 
    \right)}_{\displaystyle \coloneqq S_j^-}
\end{align}
Then, it is easy to see that 
\begin{align}
    \mathbf{B}_\delta(v) \cap \partial \mathbf{B}_\delta(u_j) \subseteq \bigcup_{i \in I_j} \Ball{\delta}{u_i}
    \Longleftrightarrow
    S_j^+ \subseteq \bigcup_{i \in I_j} \Ball{\delta}{u_i}  \quad \text{and} \quad S_j^- \subseteq \bigcup_{i \in I_j} \Ball{\delta}{u_i}.
\end{align}
We consider the covering verification problem for $S_j^+$ and $S_j^-$. 
By some linear-algebraic calculations and the continuity of $\pi$, we see that the sets $S_j^+$ and $S_j^-$ are connected. 
Therefore, if it holds that
\begin{equation}
    S_j^+ \cap \left(\bigcup_{i \in I_j} \Ball{\delta}{u_i} \right) = \bigcup_{i \in I_j} S_j^+ \cap \Ball{\delta}{u_i} \neq \emptyset,
\end{equation}
we can apply Lemma~\ref{lem: equivalence covering condition} and see that
\begin{equation}
    S_j^+ \subseteq \bigcup_{i \in I_j} \Ball{\delta}{u_i}
    \Longleftrightarrow
    \forall k \in I_{j}, \ S_j^+ \cap \partial \mathbf{B}_\delta(u_k) \subseteq \bigcup_{i \in I_{j,k}} \Ball{\delta}{u_i},
\end{equation}
where $I_{j,k} \coloneqq I_{j} \setminus \{k\}$. 
The same holds for $S_j^-$. 
Moreover, for each $i \in I_j$, it is possible to check whether $S_j^+ \cap \Ball{\delta}{u_i}$ is empty, by performing some basic linear-algebraic calculations. 
Next, we consider the covering verification problem of
\begin{equation}
    S_j^+ \cap \partial \mathbf{B}_\delta(u_k) \subseteq \bigcup_{i \in I_{j,k}} \Ball{\delta}{u_i}.
\end{equation}
We can see that
\begin{align}
    S_j^+ \cap \partial \mathbf{B}_\delta(u_k)
    = 
    \underbrace{
    \pi\left(
        \left\{ x \in S^3 \ \middle| \ 
            \begin{array}{rrr}
                 v \cdot x & > & \sqrt{1 - \delta^2}  \\
                 u_j \cdot x & = & \sqrt{1 - \delta^2} \\
                 u_k \cdot x & = & \sqrt{1 - \delta^2}
            \end{array}
        \right\}
    \right)
    }_{\displaystyle \coloneqq S_{j,k}^{++} }
    \cap \
    \underbrace{
    \pi\left(
        \left\{ x \in S^3 \ \middle| \ 
            \begin{array}{rrr}
                 v \cdot x & > & \sqrt{1 - \delta^2}  \\
                 u_j \cdot x & = & \sqrt{1 - \delta^2} \\
                 u_k \cdot x & = & -\sqrt{1 - \delta^2}
            \end{array}
        \right\}
    \right)}_{\displaystyle \coloneqq S_{j,k}^{+-} }.
\end{align}
Then, it is easy to see that
\begin{equation}
    S_j^+ \cap \partial \mathbf{B}_\delta(u_k) \subseteq \bigcup_{i \in I_{j,k}} \Ball{\delta}{u_i} 
    \Longleftrightarrow
    S_{j,k}^{++} \subseteq \bigcup_{i \in I_{j,k}} \Ball{\delta}{u_i} \quad \text{and} \quad S_{j,k}^{+-} \subseteq \bigcup_{i \in I_{j,k}} \Ball{\delta}{u_i}.
\end{equation}
We consider the covering verification problem for $S_{j,k}^{++}$ and $S_{j,k}^{+-}$. 
By some linear-algebraic calculations and the continuity of $\pi$, we see that the sets $S_{j,k}^{++}$ and $S_{j,k}^{+-}$ are connected. 
Therefore, if it holds that
\begin{equation}
    S_{j,k}^{++} \cap \left( \bigcup_{i \in I_{j,k}} \Ball{\delta}{u_i} \right)
    = 
    \bigcup_{i \in I_{j,k}} S_{j,k}^{++} \cap \Ball{\delta}{u_i} \neq \emptyset,
\end{equation}
we can apply Lemma~\ref{lem: equivalence covering condition} ans see that
\begin{equation}
    S_{j,k}^{++} \subseteq \bigcup_{i \in I_{j,k}} \Ball{\delta}{u_i}
    \Longleftrightarrow
    \forall l \in I_{j,k}, \ S_{j,k}^{++} \cap \partial \Ball{\delta}{u_l} \subseteq \bigcup_{i \in I_{j,k,l}} \Ball{\delta}{u_i},
\end{equation}
where $I_{j,k,l} = I_{j,k} \setminus \{l\}$. 
The same holds for $S_{j,k}^{+-}$. 
Moreover, for each $i \in I_{j,k}$, it is possible to check whether $S_{j,k}^{++} \cap \Ball{\delta}{u_i}$ is empty, by performing some basic linear-algebraic calculations. 
Next, we consider the covering verification problem of 
\begin{equation}
     S_{j,k}^{++} \cap \partial \Ball{\delta}{u_l} \subseteq \bigcup_{i \in I_{j,k,l}} \Ball{\delta}{u_i}. 
\end{equation}
We can see that
\begin{equation}
    S_{j,k}^{++} \cap \partial \Ball{\delta}{u_l} 
    = 
    \underbrace{
    \pi\left(
        \left\{ x \in S^3 \ \middle| \ 
            \begin{array}{rrr}
                 v \cdot x & > & \sqrt{1 - \delta^2}  \\
                 u_j \cdot x & = & \sqrt{1 - \delta^2} \\
                 u_k \cdot x & = & \sqrt{1 - \delta^2} \\
                 u_l \cdot x & = & \sqrt{1 - \delta^2}
            \end{array}
        \right\}
    \right)
    }_{\displaystyle \coloneqq S_{j,k,l}^{+++} }
    \cap \
    \underbrace{
    \pi\left(
        \left\{ x \in S^3 \ \middle| \ 
            \begin{array}{rrr}
                 v \cdot x & > & \sqrt{1 - \delta^2}  \\
                 u_j \cdot x & = & \sqrt{1 - \delta^2} \\
                 u_k \cdot x & = & \sqrt{1 - \delta^2} \\
                 u_l \cdot x & = & -\sqrt{1 - \delta^2}
            \end{array}
        \right\}
    \right)}_{\displaystyle \coloneqq S_{j,k,l}^{++-} }.
\end{equation}
Then, we show that the sets $S_{j,k,l}^{+++}$ and $S_{j,k,l}^{++-}$ are finite and contain at most two elements. 
Since the argument for $S_{j,k,l}^{+++}$ applies directly to $S_{j,k,l}^{++-}$, we focus only on $S_{j,k,l}^{+++}$. 
\begin{itemize}
    \item If ${u_j, u_k, u_l}$ are linearly independent, then the solution set of the linear equation
    \begin{equation}
    \label{eq: linear equation of u_j u_k u_l}
        \left\{ \begin{array}{rr}
            u_j \cdot x & = \sqrt{1 - \delta^2} \\
            u_k \cdot x & = \sqrt{1 - \delta^2} \\
            u_l \cdot x & = \sqrt{1 - \delta^2}
        \end{array} \right.
    \end{equation}
    is a one-dimensional affine subspace. 
    This suggests that $S_{j,k,l}^{+++}$ contains at most two elements.

    \item If ${u_j, u_k, u_l}$ are linearly dependent, from the assumption that ${u_j, u_k}$ are linearly independent, there exist unique real numbers $\alpha$ and $\beta$ such that
    \begin{equation}
        \alpha u_j + \beta u_k - u_l = 0.
    \end{equation}
    If there exists a solution $x \in \R^4$ to the linear equation~\eqref{eq: linear equation of u_j u_k u_l}, it holds that
    \begin{equation}
        (\alpha u_j + \beta u_k - u_l) \cdot x = (\alpha + \beta - 1) \sqrt{1 - \delta^2} = 0. 
    \end{equation}
    From the assumption that $\delta \in (0, \frac12]$, it is necessary that $\alpha + \beta = 1$ holds. 
    However, this contradicts the fact that 
    \begin{equation}
        \|u_l\|^2 = \| \alpha u_j + \beta u_k \|^2 = \alpha^2 + \beta^2 + 2\alpha\beta(u_j \cdot u_k) = 1
    \end{equation}
    and $u_j \cdot u_k \neq 1$ from the assumption. 
    Therefore, the linear equation~\eqref{eq: linear equation of u_j u_k u_l} has no solution, and $S_{j,k,l}^{+++}$ is empty set. 
\end{itemize}
Consequently, to solve the covering verification problem of 
\begin{equation}
    S_{j,k,l}^{+++} \subseteq \bigcup_{i \in I_{j,k,l}} \Ball{\delta}{u_i},
\end{equation}
it suffices to check the membership of at most two points in each $\Ball{\delta}{u_i}$ for $i \in I_{j,k,l}$. 
From the above discussion, the original covering verification problem can be reduced to $O(|I|^3)$ membership verification problems together with some basic linear-algebraic calculations.

\end{document}